\documentclass[prd,showpacs,floatfix,nofootinbib,twocolumn,amsmath,amssymb]{revtex4}
\usepackage{graphicx,graphics,color,dcolumn,booktabs,bm}
\usepackage{longtable,lscape}
\pdfoutput=1
\usepackage{txfonts}
\usepackage{overpic}
\usepackage{amssymb}
\usepackage{indentfirst}
\usepackage{epsfig}
\usepackage{feynmf}   
\usepackage{epstopdf}   
\usepackage{slashed}  
\usepackage{cases}
\usepackage{color}
\usepackage{multirow}

\usepackage[colorlinks, citecolor=blue,anchorcolor=red,menucolor=red, linkcolor=red,filecolor=red,runcolor=red,urlcolor=blue,frenchlinks=red]{hyperref}
\begin{document}
\title{Pseudotensor meson family}
\author{Bo Wang$^{1,2}$}\email{wangb13@lzu.edu.cn}
\author{Cheng-Qun Pang$^{1,2}$}\email{pangchq13@lzu.edu.cn}
\author{Xiang Liu$^{1,2}$\footnote{Corresponding author.}}\email{xiangliu@lzu.edu.cn}
\author{Takayuki Matsuki$^{3,4}$}\email{matsuki@tokyo-kasei.ac.jp}
\affiliation{$^1$School of Physical Science and Technology, Lanzhou University,
Lanzhou 730000, China\\
$^2$Research Center for Hadron and CSR Physics,
Lanzhou University $\&$ Institute of Modern Physics of CAS,
Lanzhou 730000, China\\
$^3$Tokyo Kasei University, 1-18-1 Kaga, Itabashi, Tokyo 173-8602, Japan\\
$^4$Theoretical Research Division, Nishina Center, RIKEN, Saitama 351-0198, Japan}
\begin{abstract}

The pseudoscalar tensor states, $\pi_2$, $\eta_2$, and $K_2$, are systematically studied through the Okubo-Zweig-Iizuka- allowed two-body strong decays, {including both the observed states reported by the Particle Data Group and the predicted states}. Phenomenological analysis combined with the experimental data {not only can} test the assignments to these discussed states, but {it can} also predict more abundant information {on} their partial decay widths, which {is helpful in the experimental study of} these observed and predicted pseudotensor states.
\end{abstract}

\maketitle

\section{INTRODUCTION}\label{section1}


Checking the observed states collected {by the} Particle Data Group (PDG) \cite{Beringer:1900zz}, we find that there is abundant experimental information about pseudotensor states with {spin parity} $J^P=2^-$, which include five $\pi_{2}$, four $\eta_{2}$, and three $K_{2}$ states.
The resonance parameters of these states are listed in Table \ref{ExpTensor}.

\renewcommand{\arraystretch}{1.3}
\begin{table}[htbp]
\caption{The resonance parameters of the observed $\pi_{2}$, $\eta_{2}$ and $K_{2}$ states.
Here, the masses and widths are average values taken
from PDG~\cite{Beringer:1900zz} and the
states listed as "further states" in PDG are marked by a superscript
$\sharp$.   \label{ExpTensor}}
\begin{center}
\begin{tabular}{cccc}
\toprule[1pt]
\midrule[1pt]
Isospin&State&                              Mass (MeV)&                          Width (MeV)  \\
\midrule[1pt]
\multirow{5}{*}{1}& $\pi_{2}(1670)$&                    $1672.2\pm 3.0$&              $260\pm9$\\

& $\pi_{2}(1880)$&                   $1895\pm 16$&                 $235\pm34$\\

& $\pi_{2}(2100)$&                   $2090\pm 29$&                 $625\pm50$\\

 &$\pi_{2}(2005)^{\sharp}$ \cite{Anisovich:2001pn}&          $2005\pm 15$&                $200\pm40$\\

 &$\pi_{2}(2285)^{\sharp}$ \cite{Anisovich:2010nh}&          $2285\pm 20\pm 25$&          $250\pm20\pm25$\\
\midrule[1pt]

\multirow{4}{*}{0}&$\eta_{2}(1645)$&                   $1617\pm5$&                     $181\pm11$\\
&$\eta_{2}(1870)$&                   $1842\pm8$&                     $225\pm14$\\
&$\eta_{2}(2030)^{\sharp}$ \cite{Anisovich:2000mv}&          $2030\pm5\pm15$&                $205\pm10\pm15$\\
&$\eta_{2}(2250)^{\sharp}$ \cite{Anisovich:2000us}&          $2248\pm20$&                     $280\pm20$\\
\midrule[1pt]
\multirow{3}{*}{$\frac{1}{2}$}&$K_{2}(1770)$&                       $1773\pm8$&                     $186\pm14$\\
&$K_{2}(1820)$&                       $1816\pm13$&                     $276\pm35$\\
&$K_{2}(2250)$&                      $2247\pm17$&                      $180\pm30$\\
\bottomrule[1pt]
\bottomrule[1pt]
\end{tabular}
\end{center}
\end{table}

Although so many pseudotensor states were observed, their underlying properties are still unknown, which is due to the absence of a systematical study of these pseudotensors. Considering the present research status of the pseudotensor states, in this work we systematically investigate the observed pseudotensor states. {First}, we discuss the possible radial assignments of these pseudotensor states. Next, we mainly focus on their Okubo-Zweig-Iizuka (OZI)-allowed two-body strong decays, which can also provide the information {on} total decay widths, because the behaviors of the OZI-allowed decays are relevant to their underlying structures. Comparing our numerical results with the experimental data, we can further test the corresponding radial assignments. What is more important is that the information {on} the obtained partial and total decay widths is valuable for further experimental study on these states. Our results { will certainly be helpful in establishing} the pseudotensor meson family.

This paper is organized as follows.
In {Sec.} \ref{section2}, we discuss how to categorize the observed states into pseudotensor meson families with {the}  help of the analysis of the Regge trajectories. In {Sec.} \ref{section3}, we perform the study of the OZI-allowed two-body strong decay of the discussed states, where the quark pair creation (QPC) model adopted in this work is briefly introduced. {Using} the phenomenological investigation by combining our results with the experimental data, we test former assignments of pseudoscalar states in {Sec.} \ref{section4}. Section \ref{section5} is devoted to the conclusions and discussion.


\section{Analysis of Regge trajectories}\label{section2}


\begin{figure*}[htbp]
\centering
\scalebox{1.5}{\includegraphics[width=\columnwidth]{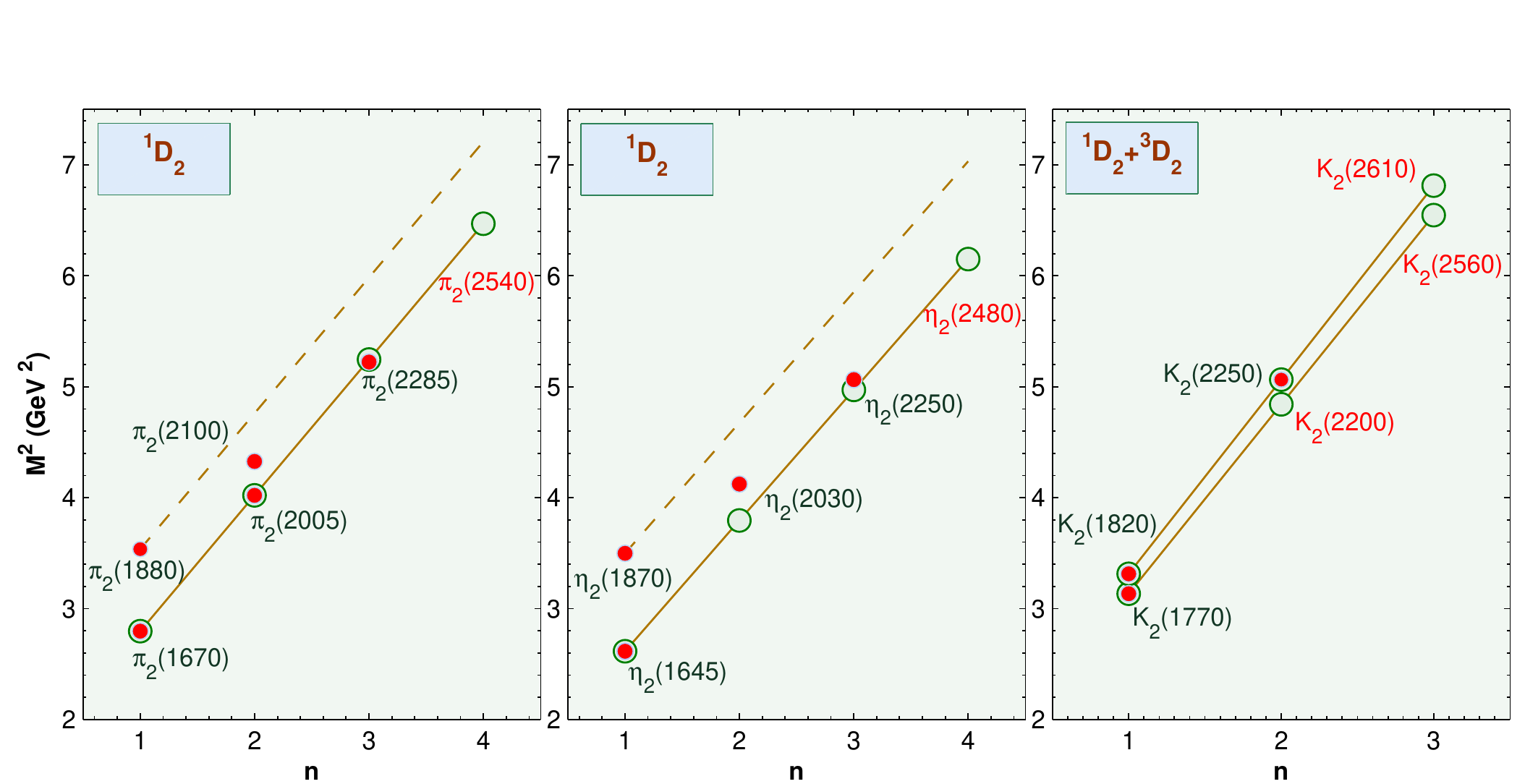}}
\caption{(color online). The analysis of the Regge trajectories for the $\pi_{2}$, $\eta_{2}$, and $K_{2}$
states. Slopes of the trajectories are 1.22, 1.18 and 1.71/1.75
GeV$^2$ for the $\pi_{2}$, $\eta_{2}$, and $K_{2}$ states, respectively. Here, {\color{green}$\bigcirc$} and {\color{red}$\bullet$} denote the theoretical and experimental values, respectively. In addition, the meson names written in red are the states still absent in experiment, where we predict their masses via the analysis of the Regge trajectories. In Ref. \cite{Bugg:2004xu}, Bugg also presented Regge trajectories with the average slope (see Fig. 1(c) in Ref. \cite{Bugg:2004xu} for more details), which is slightly different from our present analysis. \label{ReggeTraj}}
\end{figure*}

\begin{figure*}[htb]
\scalebox{1.5}{\includegraphics[width=\columnwidth]{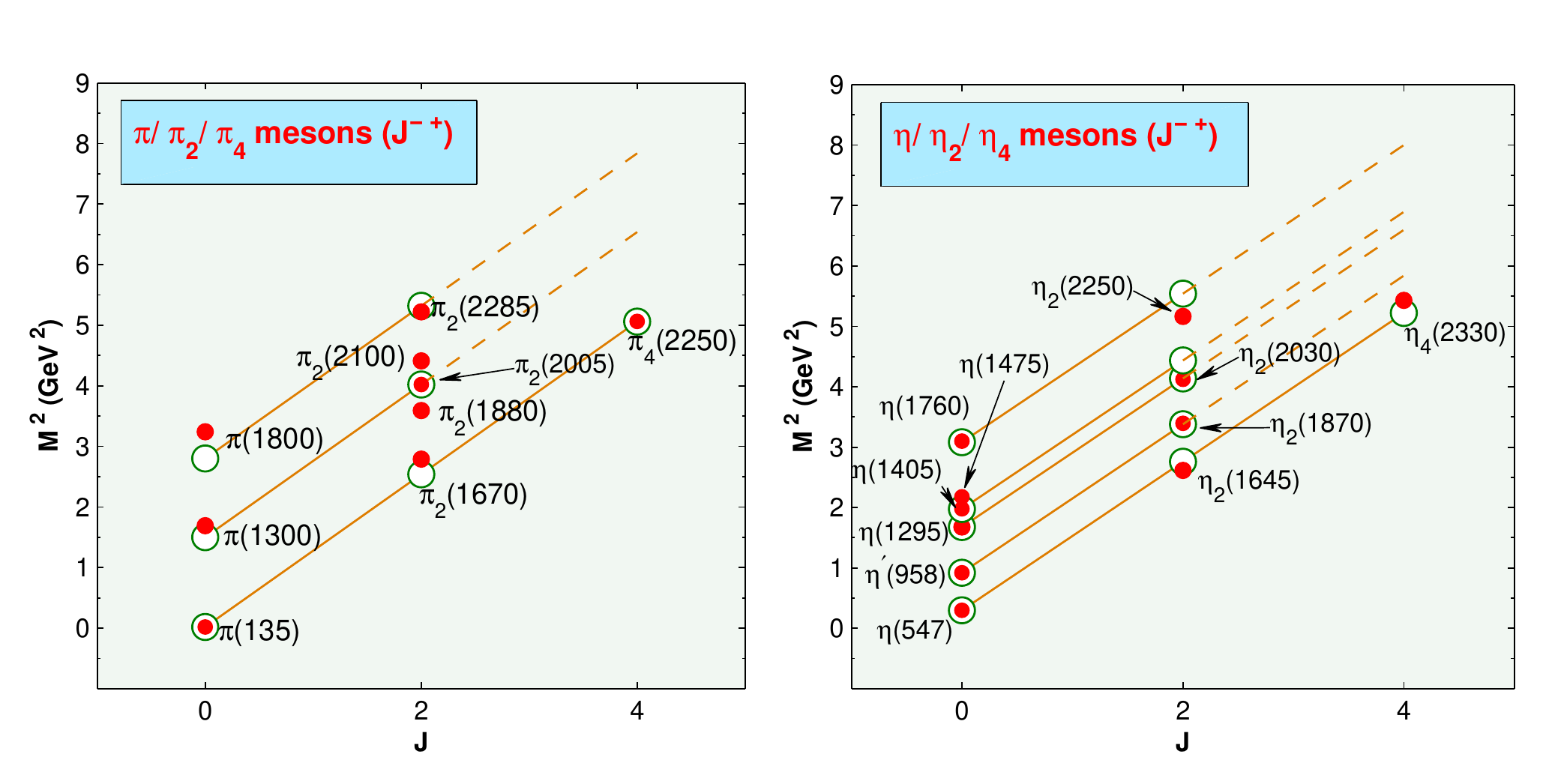}}
\caption{(color online). {The plots in $(J, M^2)$ for the $\pi/\pi_2/\pi_4$ and $\eta/\eta_2/\eta_4$ mesons. Here, {\color{green}$\bigcirc$} and {\color{red}$\bullet$} are the theoretical and experimental values, respectively. } \label{EtaJM2}}
\end{figure*}

The analysis of the Regge trajectories is an effective approach to categorize the light mesons \cite{Chew:1962eu,Anisovich:2000kxa}. In general, there is a simple relation
\begin{equation}\label{ReggeEquation}
M^{2}=M_{0}^{2}+(n-1)\mu^{2},
\end{equation}
where $M_0$ is the mass of a ground state, $M$ is the mass of a radial excitation with a radial quantum number $n$, and $\mu^2$ is {the} slope parameter of a trajectory.

{Equation} (\ref{ReggeEquation}) holds for the pseudotensor states discussed in this paper except for $\pi_{2}(1880)$ and $\eta_{2}(1870)$. In Fig. \ref{ReggeTraj}, we present the analysis of their Regge trajectories. {In addition, the plots in $(J,M^2)$ for the $\pi/\pi_2/\pi_4$ and $\eta/\eta_2/\eta_4$ mesons are also shown in Fig. \ref{EtaJM2}, which provides an extra support to the assignment listed in Fig. \ref{ReggeTraj}.}
We conclude {the following}:

(1) $\pi_{2}(1670)$, $\eta_{2}(1645)$, and $K_{2}(1770)/K_{2}(1820)$ are the ground states in the pseudotensor family. Here,
$K_{2}(1770)$ and $K_{2}(1820)$ are the mixture of the $1^1D_2$ and $1^3D_2$ states, which satisfies
\begin{equation}
\label{SpinMixingEquation1}
\left(\begin{array}{c}|K_{2}(1770)\rangle \\|K_{2}(1820)\rangle\end{array}\right)=\left(\begin{array}{cc}\cos\theta_{K(1)}&\sin\theta_{K(1)} \\-\sin\theta_{K(1)}&\cos\theta_{K(1)}\end{array}\right)\left(\begin{array}{c}|1^1D_2\rangle \\|1^3D_2\rangle\end{array}\right),
\end{equation}
where $\theta_{K(1)}$ is the corresponding mixing angle.

(2) $\pi_{2}(2005)$ [or $\pi_{2}(2100)$] and $\pi_{2}(2285)$ are the first and second radial excited states of the $\pi_{2}$ meson family, respectively. $\eta_{2}(2030)$ and $\eta_{2}(2250)$ can be the first and second radial excitations of the $\eta_2$ meson family. Additionally, $K_{2}(2200)/K_{2}(2250)$, and $K_{2}(2560)/K_{2}(2610)$, regarded as the first and second radial excitations of the $K_{2}$ meson family, have relations similar to Eq. (\ref{SpinMixingEquation1}), i.e.,
\begin{eqnarray}
\label{SpinMixingEquation2}
\left(\begin{array}{c}|K_{2}(2200)\rangle \\|K_{2}(2250)\rangle\end{array}\right)&=&\left(\begin{array}{cc}\cos\theta_{K(2)}&\sin\theta_{K(2)} \\-\sin\theta_{K(2)}&\cos\theta_{K(2)}\end{array}\right)\left(\begin{array}{c}|2^1D_2\rangle \\|2^3D_2\rangle\end{array}\right),
\end{eqnarray}
\begin{eqnarray}
\label{SpinMixingEquation3}
\left(\begin{array}{c}|K_{2}(2560)\rangle \\|K_{2}(2610)\rangle\end{array}\right)&=&\left(\begin{array}{cc}\cos\theta_{K(3)}&\sin\theta_{K(3)} \\-\sin\theta_{K(3)}&\cos\theta_{K(3)}\end{array}\right)\left(\begin{array}{c}|3^1D_2\rangle \\|3^3D_2\rangle\end{array}\right),
\end{eqnarray}
where the mixing angles $\theta_{K(2)}$ and $\theta_{K(3)}$
are introduced. We need to emphasize that $K_2(2200)$, $K_2(2560)$ and $K_2(2610)$ are predicted states (see Fig. \ref{ReggeTraj} for more details).

(3) The analysis of the Regge trajectories also indicates that it is hard to group $\pi_{2}(1880)$ into pseudotensor families which we discuss in the next section. We notice {lattice} calculations of the mass spectra of $q\bar{q}$ states and hybrids, where all obtained masses come out high because they use a value of 391 MeV for $m_\pi$. Among these predictions, the mass of a $2^{-+}$ hybrid is estimated as $\sim1880$ MeV \cite{lattice}. Thus, $\pi_{2}(1880)$ can be {a good candidate for the} $2^{-+}$ hybrid.

(4) The masses of $\pi_2$ and $\eta_2$ with $n=4$ states are predicted, {and} are named $\pi_2(2540)$ and $\eta_2(2480)$. Both of the states are still missing in experiment.

(5) {The plots in $(J, M^2)$ for the $\eta/\eta_2/\eta_4$ mesons (see Fig. \ref{EtaJM2}) show that $\eta_2(1870)$ {can} be the partner of $\eta_2(1645)$, which is similar to the relation between $\eta(547)$ and $\eta^\prime(958)$.
Later, we will discuss this possibility of $\eta_2(1870)$ as the partner of $\eta_{2}(1645)$. }

The analysis presented in { Figs. \ref{ReggeTraj}}-\ref{EtaJM2} is only a rough estimate {of} the mass spectrum of the states studied in this paper. Such categorization should be tested by further dynamical study. In Sec. \ref{section3}, we calculate their two-body OZI-allowed decays.


\begin{figure*}[htbp]
\begin{center}
\scalebox{2.3}{\includegraphics[width=\columnwidth]{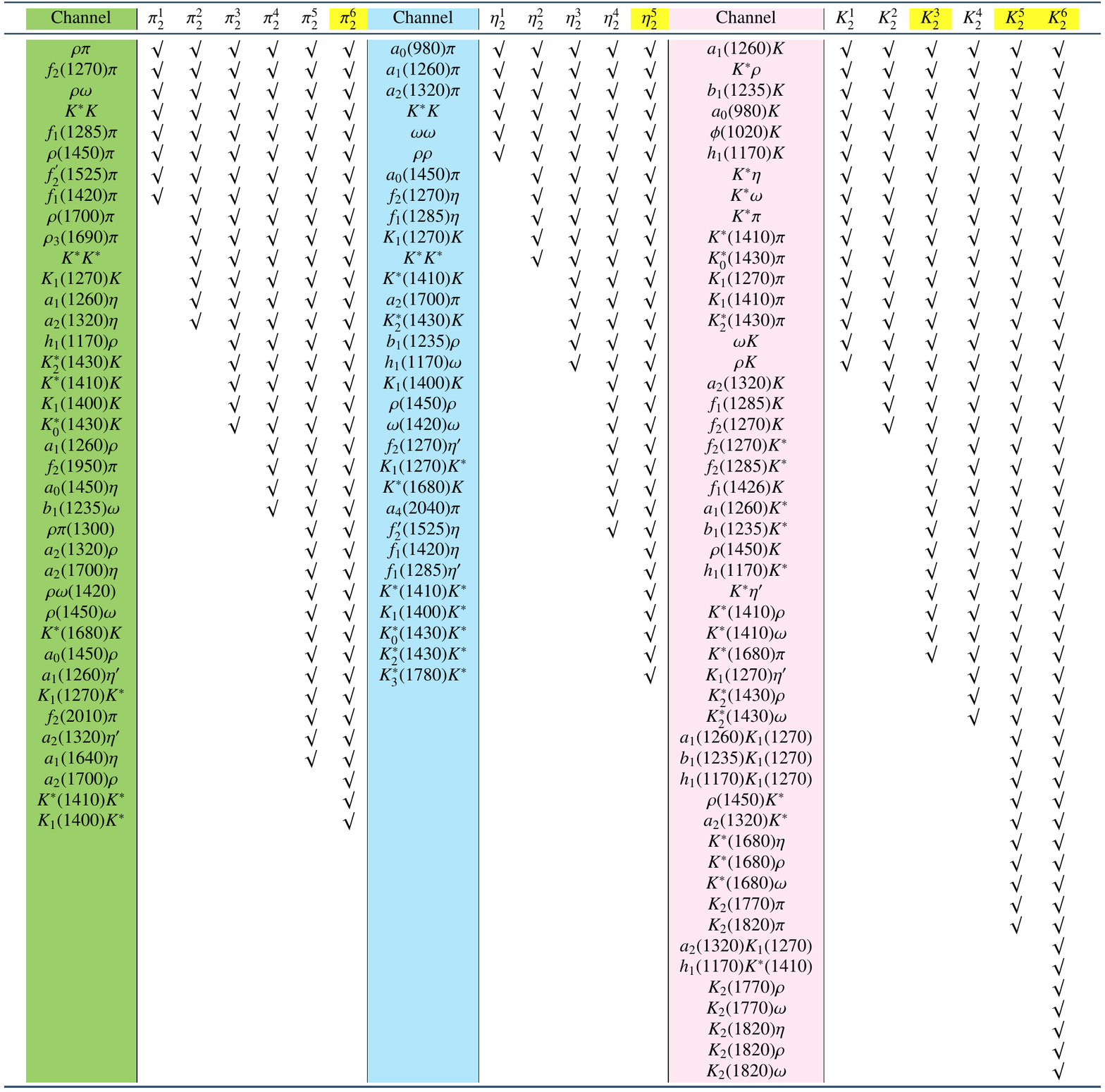}}
\caption{(color online). The OZI-allowed two-body decay modes of the $\pi_{2}$, $\eta_{2}$, and $K_{2}$ states. Here, $\omega$, $\rho$, and $\eta^\prime$ denote $\omega(782)$, $\rho(770)$, and $\eta'(958)$, respectively. The OZI-allowed two-body decays are marked by $\surd$. $\pi_2^1/\eta_2^1/K_2^1$, $\pi_2^2/\eta_2^2/K_2^2$, $\pi_2^3/\eta_2^3/K_2^3$, $\pi_2^4/\eta_2^4/K_2^4$, $\pi_2^5/\eta_2^5/K_2^5$, and $\pi_2^6/K_2^6$ correspond to
$\pi_2(1670)/\eta_2(1645)/K_2(1770)$, $\pi_2(1880)/\eta_2(1870)/K_2(1820)$, $\pi_2(2005)/\eta_2(2030)/K_2(2200)$, $\pi_2(2100)/\eta_2(2250)/K_2(2250)$, $\pi_2(2285)/\eta_2(2480)/K_2(2560)$, {and} $\pi_2(2540)/K_2(2610)$, respectively. In addition, we mark these predicted states in the first column {with} the yellow background.\label{DecayTable}}
\end{center}
\end{figure*}

\section{A BRIEF INTRODUCTION OF THE QPC MODEL}\label{section3}
There are several quark models \cite{Hayot:1973by,LeYaouanc:1974ri,Le Yaouanc:1974mr,Eichten:1978tg,Dosch:1986dp} to deal with the strong decay of hadrons, { and among these} the QPC model is one of the most popular. In 1968, Micu proposed the QPC model in Ref. \cite{Micu:1968mk}, { and  then it} was further developed by the Orsay {Group}. Later, the QPC model {was} widely applied to calculate the OZI-allowed two-body strong decays of hadrons \cite{vanBeveren:1979bd,vanBeveren:1982qb,Bonnaz:2001aj,Lu:2006ry,Luo:2009wu,Blundell:1996as,Page:1995rh,Capstick:1993kb,Capstick:1986bm,
Ackleh:1996yt,Close:2005se,Pang:2014laa,Zhou:2004mw,Guo:2005cs,Zhang:2006yj,Sun:2009tg,Liu:2009fe,Sun:2010pg,Yu:2011ta,Wang:2012wa,Ye:2012gu}.
In this model, to depict the quark-antiquark pair created from the QCD vacuum with vacuum a quantum number $J^{PC}=0^{++}$, the transition operator is introduced, i.e.,
\begin{eqnarray} \label{TransitionOperator}
\mathcal{T}&=&-3\gamma\sum_{m} \langle1m; 1-m|00\rangle \int d^3\mathbf{p}_3d^3\mathbf{p}_4\delta^3(\mathbf{p}_3+\mathbf{p}_4) \nonumber \\
           &&\times \mathcal{Y}_{1m}\left(\frac{\mathbf{p}_3-\mathbf{p}_4}{2}\right)\chi^{34}_{1,-m}\phi^{34}_0\omega^{34}_0
           b^{\dagger}_{3i}(\mathbf{p}_3)d^{\dagger}_{4i}(\mathbf{p}_4).
\end{eqnarray}
\par
In the above expression, $\gamma$ is a dimensionless parameter to describe the strength of the $q\bar{q}$ pair creation, which can be obtained by fitting the experimental data systematically. {In} numerical calculations we set $\gamma=8.7$ for the $u \bar u$ (or $d \bar d$) pair creation ({see Table II in Ref. \cite{Ye:2012gu} for more details about extracting the $\gamma$ value}), while for the strength of the $s\bar s$, we take $\gamma=8.7/\sqrt{3}$ \cite{LeYaouanc:1977gm}. Here, $\mathbf{p}_3(\mathbf{p}_4)$ denotes the three-momentum of a quark (an antiquark) created from the vacuum. Then, the transition matrix element for the process of $A\to B+C$  can be expressed as
\begin{eqnarray}
\langle
BC|\mathcal{T}|A\rangle=\delta^3(\mathbf{P}_B+\mathbf{P}_C)\mathcal{M}^{M_{J_A}M_{J_B}M_{J_C}},\label{matrix}
\end{eqnarray}
where the magnetic momentum for the decay meson is denoted by $M_{J_{i}}\ (i=A,B,C)$, $\mathbf{P}_B(\mathbf{P}_C)$ is the three-momentum of the final particle $B(C)$ in the rest frame of the initial state $A$, and  {{$\mathcal{M}^{M_{J_A}M_{J_B}M_{J_C}}$ denotes the calculated amplitude
 }. We mark the created quark and antiquark with the subscripts 3 and 4, {respectively}, in Eq. (\ref{TransitionOperator}) . $\chi$ is the spin wave function and $\chi^{34}_{1,-m}$ corresponds to a spin triplet notation, where $i$ is the $SU(3)$ color indices of the
quark-antiquark pair created from the vacuum with $J^{PC}=0^{++}$. $\phi$ and $\omega$ denote the flavor and color wave functions, respectively, i.e., $\phi^{34}_0=(u \bar u +d \bar d +
s\bar s)/\sqrt{3}$ and $\omega^{34}_0=\delta_{\alpha_3\alpha_4}/\sqrt{3}$
$(\alpha_i=1,2,3)$. Additionally, $\mathcal{Y}_{\ell m}(\mathbf{p})\equiv
|\mathbf{p}|^\ell Y_{\ell m}(\theta_p,\phi_p)$ is the $\ell${th}
solid harmonic polynomial.

{Finally, the general two-body decay width can be represented as
\begin{eqnarray}
\Gamma_{A\rightarrow BC}=\pi^2\frac{|\mathbf{P}_B|}{m_A^2}\sum_{J,L}\left|\mathcal{M}^{JL}(A\rightarrow
BC)\right|^2
\end{eqnarray}
with
\begin{eqnarray}
&&\mathcal{M}^{JL}(A\rightarrow
BC)\nonumber\\&&=\frac{\sqrt{2L+1}}{2J_A+1}\sum_{M_{J_B},M_{J_C}}\langle
L0;JM_{J_A}|J_A M_{J_A}\rangle\nonumber \\ \nonumber && \quad \times
\langle J_B M_{J_B};J_C M_{J_C}|J M_{J_A}\rangle
\mathcal{M}^{M_{J_A}M_{J_B}M_{J_C}}, \nonumber\\
\label{hhh}
\end{eqnarray}
which is obtained by using the Jacob-Wick formula \cite{Jacob:1959at,Chung:1971ri}. {{For {the} readers' convenience,  we add a detailed deduction of Eq. (\ref{hhh}) in {the} Appendix.}}
In the above expressions, $m_A$ is the mass of an initial particle $A$. For the concrete calculation by the QPC model, a simple harmonic oscillator (SHO) wave function is adopted to describe the spacial wave function of a meson, which has the form \footnote{{{Here, in the momentum space the SHO wave function is expressed as
\begin{eqnarray*}
\Psi_{n,\ell,m}(R,\mathbf{q})&=&{(-1)^{n-1}(-i)^\ell}{R^{3/2}}\sqrt{\frac{2(n-1)!}{\Gamma(n+\ell+1/2)}}\left(q R\right)^\ell
e^{\frac{-q^2 R^2}{2}} \nonumber \\ &&\times L_{n-1}^{\ell+1/2}(q^2 R^2)Y_{\ell m}(\theta_q,\phi_q),\label{k11}
\end{eqnarray*}
where $L_{n-1}^{\ell+1/2}(q^2 R^2)$ is an associated Laguerre polynomial,\ and $R$ is an oscillator parameter.}}}
\begin{eqnarray}
\Psi_{n,\ell,m}(R,\mathbf{q})&=&\mathcal{R}_{n,\ell}(R,\mathbf{q})\mathcal{Y}_{\ell m}(\mathbf{q})\nonumber\\
&=&\mathcal{N}_{n,\ell}\exp\left(-\frac{R^2 q^2}{2}\right)|\mathbf{q}|^\ell Y_{\ell m}(\theta_q,\phi_q)\mathcal{P}(\mathbf{q}^2).
\end{eqnarray}
Here, $\mathcal{N}_{n,\ell}$ represents a normalization coefficient  and
$\mathcal{P}(\mathbf{q}^2)$ denotes a polynomial in terms of $\mathbf{q}^2$.} In Ref. \cite{Luo:2009wu}, the authors once gave a detailed review of the QPC model and the calculation of the  transition amplitude $\langle BC|\mathcal{T}|A\rangle$. Thus, the reader can consult Ref.
\cite{Luo:2009wu} for more details.
In addition, we need to explain how to constrain the $R$ value in the SHO wave function. Usually, $R$ can be obtained such that it reproduces the realistic root mean square radius which is determined by solving the Schr\"{o}dinger equation with the potential given in Ref. \cite{Close:2005se}.

The allowed two-body strong decay modes of $\pi_2/\eta_2/K_2$ states are listed in Fig. \ref{DecayTable}. We obtain their partial and total decay widths via the QPC model. In the next section, we perform a phenomenological analysis by comparing our theoretical results with the experimental information, which will be helpful and meaningful for future experiments to comprehensively understand the underlying properties of these $\pi_{2}/\eta_{2}/K_{2}$ states.


\section{PHENOMENOLOGICAL ANALYSIS}\label{section4}

With the above preparation, in the following we carry out the analysis by combining our results with the experimental data, which can be applied to test whether the assignment discussed in Sec. \ref{section2} is reasonable. Before illustrating the concrete analysis for each meson family, we briefly review the corresponding experimental and theoretical research status.

\subsection{$\pi_{2}$ meson family}

$\pi_{2}(1670)$ was first reported in Ref. \cite{Bartsch:1968zz} in the reaction $\pi^{+}p\to p\pi^+\pi^+\pi^-$. In 1968, Baltay {\it et al.} observed a negative G-parity state at 1630 MeV \cite{Baltay:1968zza}, which was confirmed in Ref. \cite{Ascoli:1973nj} with the mass and width, $M=1660\pm10$ MeV and $\Gamma=270\pm60$ MeV, respectively. By the double $\gamma$ scattering experiments, the CELLO and Crystal Ball Collaborations observed $\pi_{2}(1670)$ in the reactions $\gamma\gamma\to \pi^{0}\pi^{0}\pi^{0}$ and $\gamma\gamma\to \pi^{+}\pi^{-}\pi^{0}$ \cite{Behrend:1989rc,Antreasyan:1990zm}. In 1998, the WA102 Collaboration reported the $J^{PC}=2^{-+}$ state interacting with $\rho^{\pm}\pi^{\mp}$ via a $P$-wave and $f_{2}(1270)\pi^{0}$ via an $S$-wave in the reaction of $pp\to p_{f}(\pi^{+}\pi^{-}\pi^{0})p_{s}$ \cite{Barberis:1998in}. The E852 experiment {performed} the partial wave analysis of the reaction $\pi^{-}p\to \pi^{+}\pi^{-}\pi^{-}p$ and confirmed $\pi_{2}(1670)$, which strongly decays into $\rho\pi$ via a $P$-wave and $f_2(1270)\pi$ via an $S$-wave \cite{Chung:2002pu}. In 2005, its main decay mode $\rho\omega$ was observed by the E852 Collaboration in the process of $\pi^{-}p\to \omega\pi^{-}\pi^{0}p$, where there also exists the evidence of $\pi_{2}(1880)$ and $\pi_{2}(2005)$ \cite{Lu:2004yn}. Four years ago, the COMPASS Collaboration also reported the same structure in the $f_{2}(1270)\pi$ channel in the reaction $\pi^{-}Pb\to \pi^{-}\pi^{-}\pi^{+}Pb^{\prime}$ \cite{Alekseev:2009aa}. By the above experimental efforts, $\pi_{2}(1670)$ was experimentally established. At present, the average mass and width of $\pi_{2}(1670)$ listed in PDG \cite{Beringer:1900zz} are $1672.2 \pm 3.0$ and $260 \pm 9$ MeV, respectively.

In the double $\gamma$ scattering reaction, an enhancement near 1.8 GeV was also reported \cite{Behrend:1989rc,Antreasyan:1990zm}, which is referred to {as} $\pi_{2}(1880)$.  A similar structure to $J^{P}=2^{-}$ was given by the VES Collaboration subsequently in the $a_{2}(1320)\eta$ channel in the collected $\eta\eta\pi^{-}$ data \cite{Amelin:1995fg}. The Crystal Barrel Collaboration \cite{Adomeit:1996nr} analyzed the data of $p\bar{p}\to \eta\eta\pi^{0}\pi^{0}$, which indicates the existence of a resonance decaying strongly into $a_{2}(1320)\eta$ but weakly into $f_{0}(1500)\pi$ with the mass and width $M=1880\pm20$ MeV and $\Gamma=255\pm45$ MeV, respectively  \cite{Anisovich:2001hj}. This state was also confirmed by E852 in the $f_1(1285)\pi$  \cite{Kuhn:2004en} and $\rho\omega$ channels \cite{Lu:2004yn}. In 2008, the E852 Collaboration observed a signal for $\pi_{2}(1880)$ in the $a_2(1320)\eta$ channel associated with $\pi_{2}(1670)$ \cite{Eugenio:2008zza}. The decay behaviors of $\pi_{2}(1880)$ strongly coupling with the $a_{2}(1320)\eta$ channel makes $\pi_{2}(1880)$  an isotriplet partner of $\eta_{2}(1870)$, which dominantly decays into $f_{2}(1270)\eta$ and $a_{2}(1320)\pi$ \cite{Anisovich:2000mv}.

$\pi_{2}(1880)$ is the most controversial meson in the observed $\pi_{2}$ states since the mass is too light to be the first radial excitation of $\pi_{2}(1670)$. Thus, the assignment of $\pi_{2}(1880)$ to a hybrid was first proposed by Anisovich {\it et al}. \cite{Anisovich:2001hj}, which was discussed by other theoretical groups \cite{Anisovich:2001pp,Kuhn:2004en,Lu:2004yn,Eugenio:2008zza,Klempt:2007cp,Bugg:2004xu}, where a main motivation is that the mass of $\pi_{2}(1880)$  just falls into the prediction of the flux-tube model; i.e., the predicted mass is 1.8$-$1.9 GeV for a $J^{PC}=2^{-+}$ hybrid \cite{Barnes:1995hc}. Additionally, in Refs. \cite{Li:2008xy,Barnes:1996ff}, the decay behaviors of $\pi_2(1880)$
as the first radial excitation of $\pi_{2}(1670)$ or hybrid were studied, where $\pi_2(1880)$
has distinctive features under these two assignments (see Refs. \cite{Li:2008xy,Barnes:1996ff} and two reviews \cite{Klempt:2007cp,Meyer:2010ku} for the detailed discussions). Considering the above situation of $\pi_{2}(1880)$, we do not include $\pi_{2}(1880)$ in our study in this work.

In the following, we introduce $\pi_{2}(2005)$. In the partial wave analysis of $p\bar{p}\to 3\pi^{0},\ \pi^{0}\eta,\ \pi^{0}\eta^{\prime}$  \cite{Anisovich:2001pp} from the  Crystal Barrel experiment, Anisovich {\it et al.} found an evidence for the $2^{-+}$ state with the mass $M=2005\pm15$ MeV and width $\Gamma=200\pm40$ MeV.  Subsequently, a $2^{-+}$ structure with the mass $M=2003\pm88\pm148$ and width $\Gamma=306\pm132\pm121$, was revealed by the E852 Collaboration in the $f_{1}(1285)\pi$ channel of the reaction $\pi^{-}p\to \eta\pi^{+}\pi^{-}\pi^{-}p$ \cite{Kuhn:2004en}. Similarly, the reaction $\pi^{-}p\to \pi^{+}\pi^{-}\pi^{-}\pi^{0}\pi^{0}p$ measured by the E852 Collaboration shows that the $\pi_{2}(2005)$ signal appears in the $\omega\rho^{-}$ decay channel \cite{Lu:2004yn}.

In 1980, the ACCMOR Collaboration observed a $J^P=2^{-}$ resonance  with the mass $M=2100\pm150$ and width $\Gamma=651\pm50$ in the $\pi^{-}p\to \pi^{-}\pi^{-}\pi^{+}p$ process \cite{Daum:1980ay}. Here, we need to comment {on} the anomalously large width measured by ACCMOR \cite{Daum:1980ay}. In Ref. \cite{Daum:1980ay}, they missed $\pi_2(1880)$, which results in the large width they found . \footnote{We would like to thank David Bugg for the explanation on this point.} In Ref. \cite{Amelin:1995gu}, the VES Collaboration studied the $\pi^{-}A\to \pi^{+}\pi^{-}\pi^{-}A$ reaction, where there exists a structure with the mass $M=2090\pm30$, {and} width $\Gamma=520\pm100$, respectively. These observations correspond to the $\pi_2(2100)$ state listed in PDG.

In addition, the reanalysis of experimental data carried by the {Crystal Barrel} Collaboration \cite{Adomeit:1996nr} indicates that a $I=1$ and $J^{PC}=2^{-+}$ state may exist with the mass $M=2285\pm20\pm25$ and width $\Gamma=250\pm20\pm25$ \cite{Anisovich:2010nh}, which was listed in the further states of PDG as $\pi_2(2285)$.

Since the number of observed $\pi_2$ states is larger than that of the allowed $\pi_2$ mesons, in this work we study the decays of $\pi_2(1670)$, $\pi_2(2005)/\pi_2(2100)$, $\pi_2(2285)$ as the ground, the first, and the second radial excitations in the $\pi_2$ meson family. According to the analysis of the Regge trajectories, we can predict that the third radial excitation of {the} $\pi_2$ meson is 2540 MeV, which is named $\pi_2(2540)$. Its decay behavior is also predicted in this work.

\subsubsection{$\pi_{2}(1670)$}

\begin{figure}[htb]
\scalebox{0.98}{\includegraphics[width=\columnwidth]{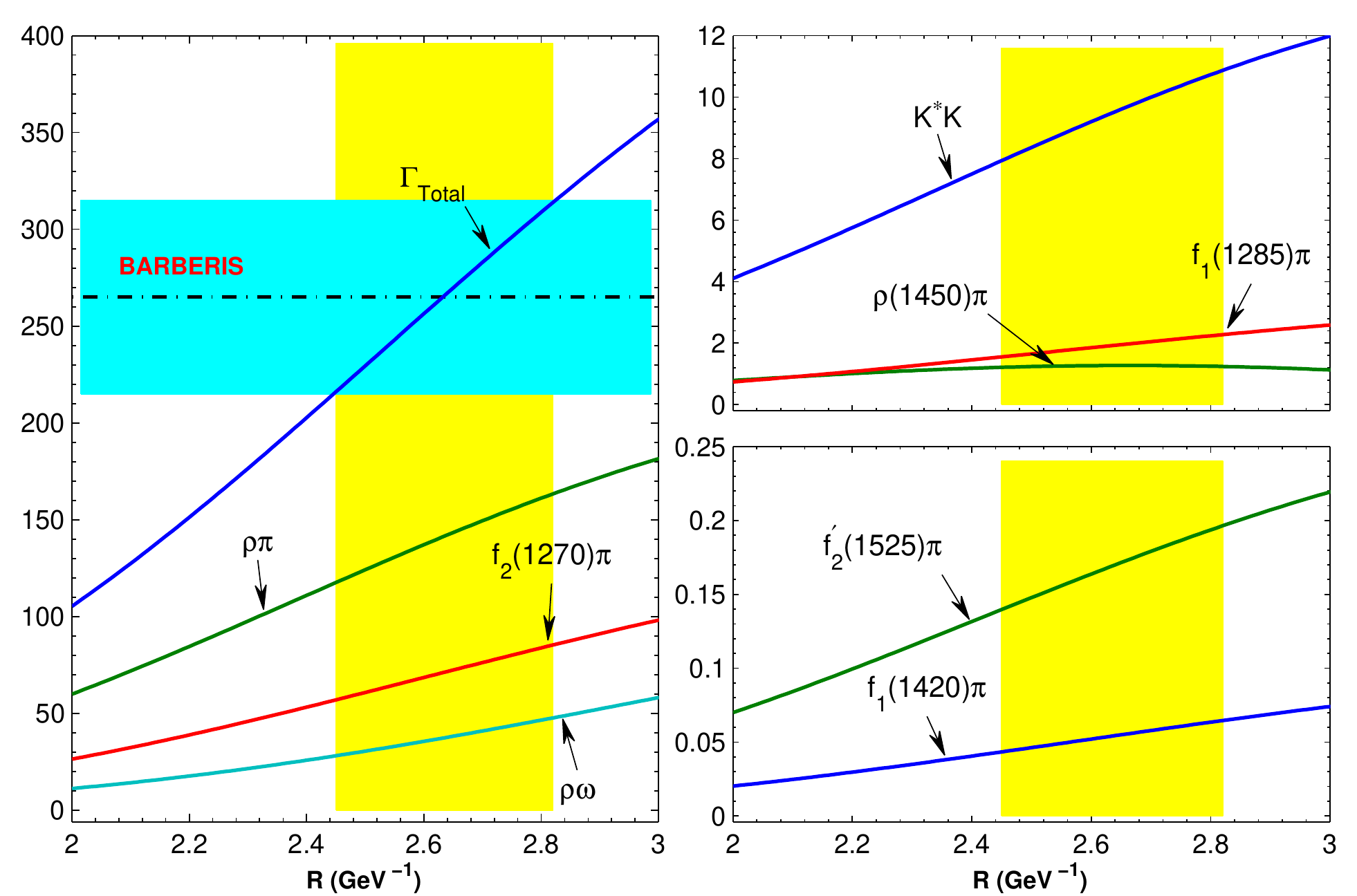}}
\caption{(color online). The $R$ dependence of {the} calculated partial and total decay widths of $\pi_{2}(1670)$. Here, the
horizontal dot-dashed line with band is the experimental total width from Ref.
\cite{Barberis:2001bs}. All results are in units of MeV. The vertical yellow bands denote the allowed $R$ value range, where the theoretical result overlaps with the experimental data in Ref. \cite{Kokoski:1985is}. \label{Pi21670}}
\end{figure}

\begin{table}[htb]
\caption{The comparison of the theoretical and experimental  values for $\pi_{2}(1670)$. Here, all theoretical values are obtained by taking the typical value $R=2.6$ GeV$^{-1}$. $\Gamma_{Total}$ denotes the total width. \label{Pi21670BrTable}}
\begin{center}
 \renewcommand{\arraystretch}{1.5}
 \tabcolsep=1.4pt
\begin{tabular}{lcc}
\toprule[1pt]
\toprule[1pt]
Ratios&                     This work&                   Experimental data \cite{Beringer:1900zz} \\
\midrule[1pt]

$\Gamma(f_{2}(1270)\pi)/\Gamma(\rho\pi)$&                $0.5$&                   $2.33\pm0.21\pm0.31$\\

$\Gamma(\bar{K^{\ast}}K+c.c)/\Gamma(f_{2}(1270)\pi)$&   $0.13$&                  $0.075\pm0.025$\\

$\Gamma(\rho\omega)/\Gamma_{Total}$&                      $0.14$&                  $0.027\pm0.004\pm0.01$\\

$\Gamma(\rho(1450)\pi)/\Gamma_{Total}$&                    $<0.005$&                 $<0.0036$\\

$\Gamma(b_{1}(1235)\pi)/\Gamma_{Total}$&                   $0$&                         $0.0019$\\

\bottomrule[1pt]
\bottomrule[1pt]
\end{tabular}
\end{center}
\end{table}

$\pi_{2}(1670)$ is a {well-established} $\pi_{2}$ meson with $1^{1}D_{2}$, which is illustrated by our calculation in Fig. \ref{Pi21670}, where the $\rho\pi$, $f_{2}(1270)\pi$, and $\rho\omega$ decay channels are dominant, which is consistent with the experimental data listed in PDG \cite{Beringer:1900zz}.
 The calculated total width is in agreement with the data given in Ref. \cite{Barberis:2001bs}.

In Table \ref{Pi21670BrTable}, we also list the calculated typical branching ratios, which are comparable with those calculated in Ref. \cite{Barnes:1996ff}. We further compare our results with the experimental data, where the obtained total decay width (265 MeV), the decay ratios $\Gamma(f_{2}(1270)\pi)/\Gamma(\rho\pi)$, $\Gamma(\bar{K^{\ast}}K+c.c)/\Gamma(f_{2}(1270)\pi)$, and  $\Gamma(\rho(1450)\pi)/\Gamma_{Total}$ are qualitatively consistent with the experimental data. For the ratio $\Gamma(b_{1}(1235)\pi)/\Gamma_{Total}$, our calculation indicates that it is zero due to the constraint of the spin selection rule. We also obtain that the branching ratio of $\pi_2(1670)\to \rho\omega$ is 0.14, which is far larger than the experimental value given in Ref. \cite{Amelin:1999gk}. Thus, these differences can be further clarified by more experimental efforts in {the} future.

Since $\pi_{2}(1670)\to\rho\pi$ can occur via {the} $P$-wave and $F$-wave, we can separately consider the $P$-wave and $F$-wave contributions to
the partial decay width of $\pi_{2}(1670)\to\rho\pi$, where we obtain $\Gamma(\pi_{2}(1670)\to\rho\pi)_P/\Gamma(\pi_{2}(1670)\to\rho\pi)_F=0.89$ (we use subscripts $P$ and $F$ to distinguish two contributions). Similarly, we also obtain $\Gamma(\pi_{2}(1670)\to f_{2}(1270)\pi)_D/\Gamma(\pi_{2}(1670)\to f_{2}(1270)\pi)_S=0.08$, where the subscripts $S$ and $D$ are adopted to mark the $S$-wave and $D$-wave contributions, respectively. These are consistent with the corresponding experimental data in Refs. \cite{Barberis:1998in,Alekseev:2009aa} which {show} that $\pi_{2}(1670)$ strongly couples to $f_{2}(1270)\pi$ via an $S$-wave. We need to specify that the above ratios are estimated by taking {the} typical value $R=2.6$ GeV$^{-1}$ \cite{Kokoski:1985is}.

\subsubsection{$\pi_2(2005)/\pi_{2}(2100)$}

By the analysis of the Regge trajectories shown in Fig. \ref{ReggeTraj}, $\pi_2(2005)$ can be the possible candidate of the first radial excitation of $\pi_2(1670)$ since its mass is in good agreement with the theoretical prediction with the slope $\mu^{2}=1.22$ GeV$^2$. However, we notice that there is another state $\pi_2(2100)$ near 2.0 GeV, which is listed in {the} meson summary table of PDG \cite{Beringer:1900zz}.  Comparing it with $\pi_2(2100)$, $\pi_2(2005)$ is treated as a further state listed in PDG \cite{Beringer:1900zz}. If only taking into account the analysis of the Regge trajectories, the mass of $\pi_{2}(2100)$ slightly deviates from the theoretical value of the first radial excitation of $\pi_2(1670)$.
Due to {the} above situation, in the following we study the decay behavior of $\pi_2(2005)$ and $\pi_2(2100)$ by combining with the corresponding experimental data, where both $\pi_2(2005)$ and $\pi_(2100)$ are considered as the first radial excitation of $\pi_2(1670)$.

\begin{figure}[htbp]
\scalebox{1.}{\includegraphics[width=\columnwidth]{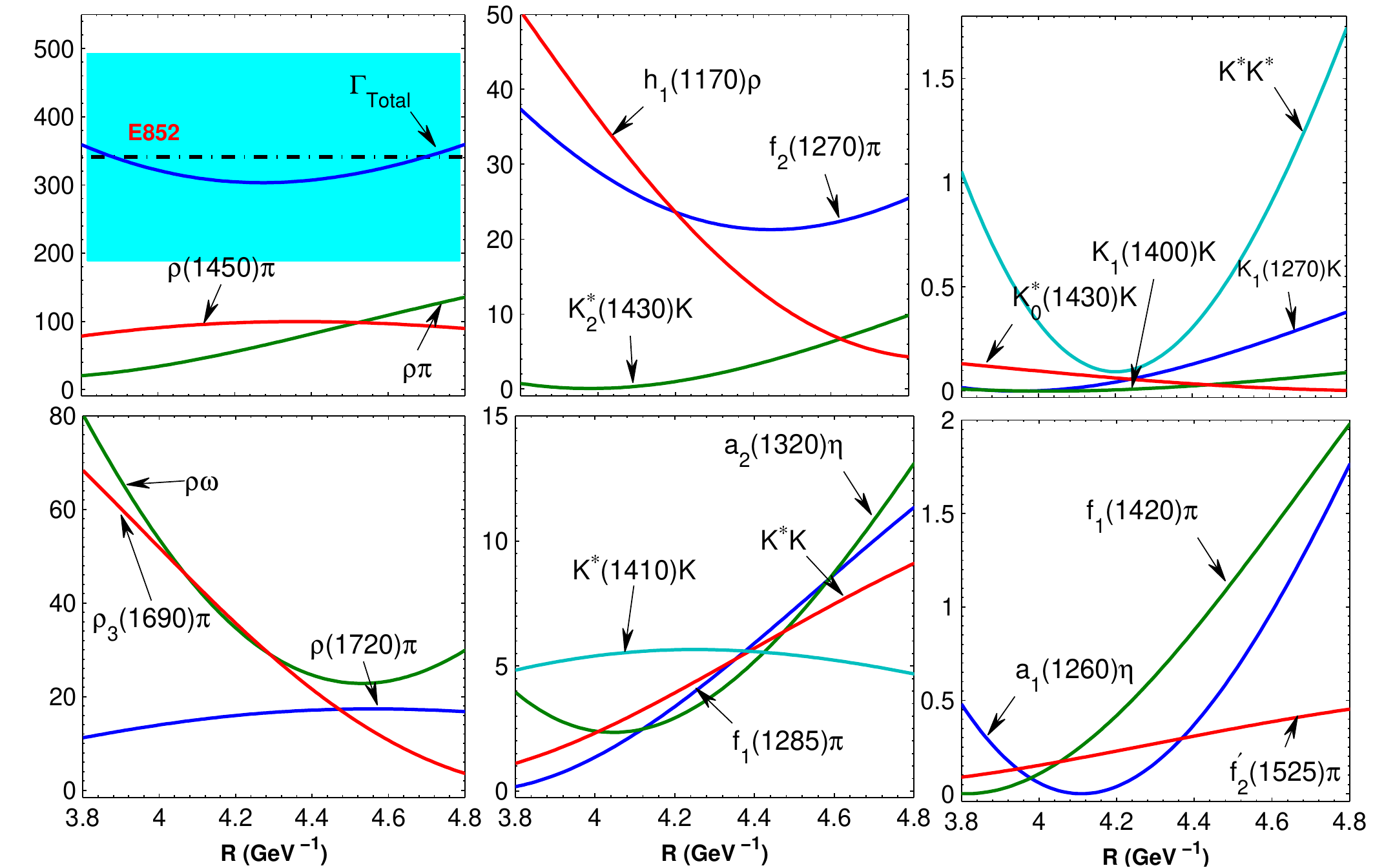}}
\caption{(color online). The $R$ dependence of {the} calculated partial and total decay widths of $\pi_{2}(2005)$. Here, the
dot-dashed line with the horizontal band is the experimental total width from Ref.
\cite{Lu:2004yn}. All obtained partial and total decay widths are in units of MeV. \label{Pi22005}}
\end{figure}

\begin{figure}[htbp]
\scalebox{0.98}{\includegraphics[width=\columnwidth]{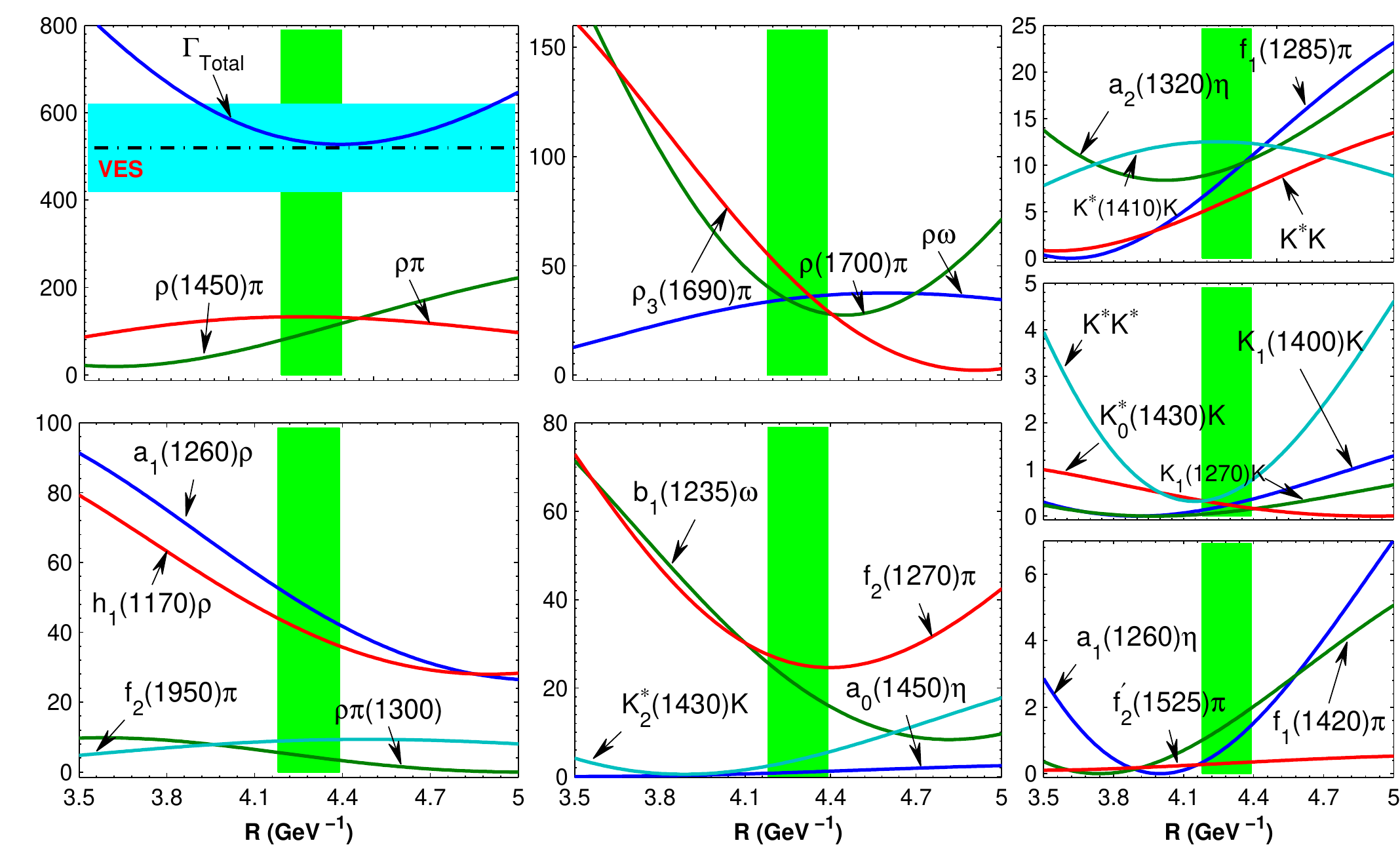}}
\caption{(color online). The $R$ dependence of {the} calculated partial and total decay widths of $\pi_{2}(2100)$. Here, the
dot-dashed line with the horizontal band is the experimental total width from Ref.
 \cite{Amelin:1995gu}. {The green} band corresponds to the range of the $R$ value, where the experimental data can be fitted with our theoretical results.
 All obtained partial and total decay widths are in units of MeV. \label{Pi22100}}
\end{figure}

\begin{figure*}[htbp]
\scalebox{1.5}{\includegraphics[width=\columnwidth]{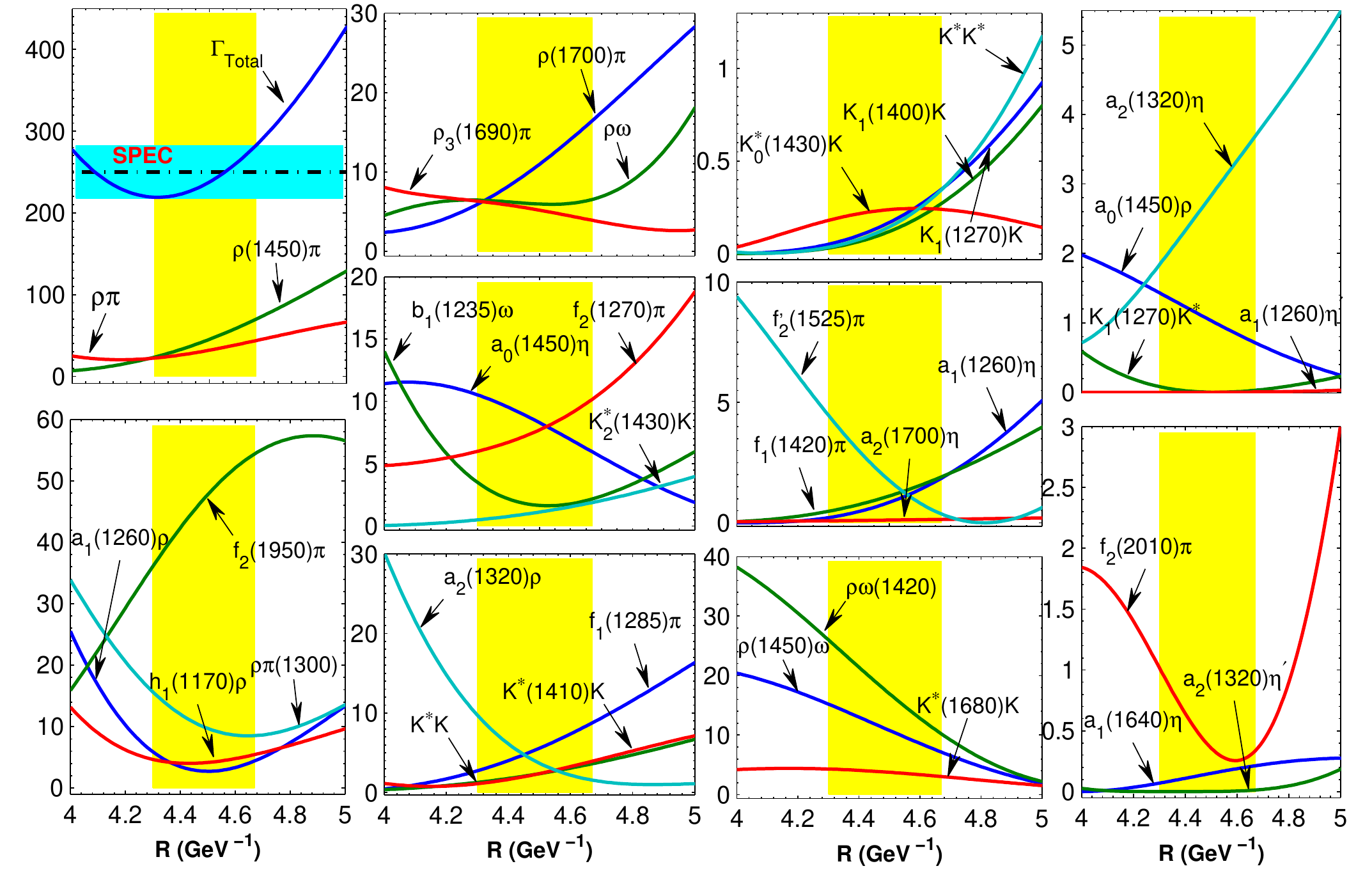}}
\caption{(color online). The $R$ dependence of the partial and total decay widths of $\pi_{2}(2285)$. Here, the dot-dashed line with the horizontal band is the experimental total width from Ref.
\cite{Anisovich:2010nh}. The vertical band denotes that the theoretical result overlaps with the experimental data
when taking the corresponding $R$ range.
All results are in units of MeV. \label{Pi22285}}
\end{figure*}

\begin{figure*}[htbp]
\scalebox{1.5}{\includegraphics[width=\columnwidth]{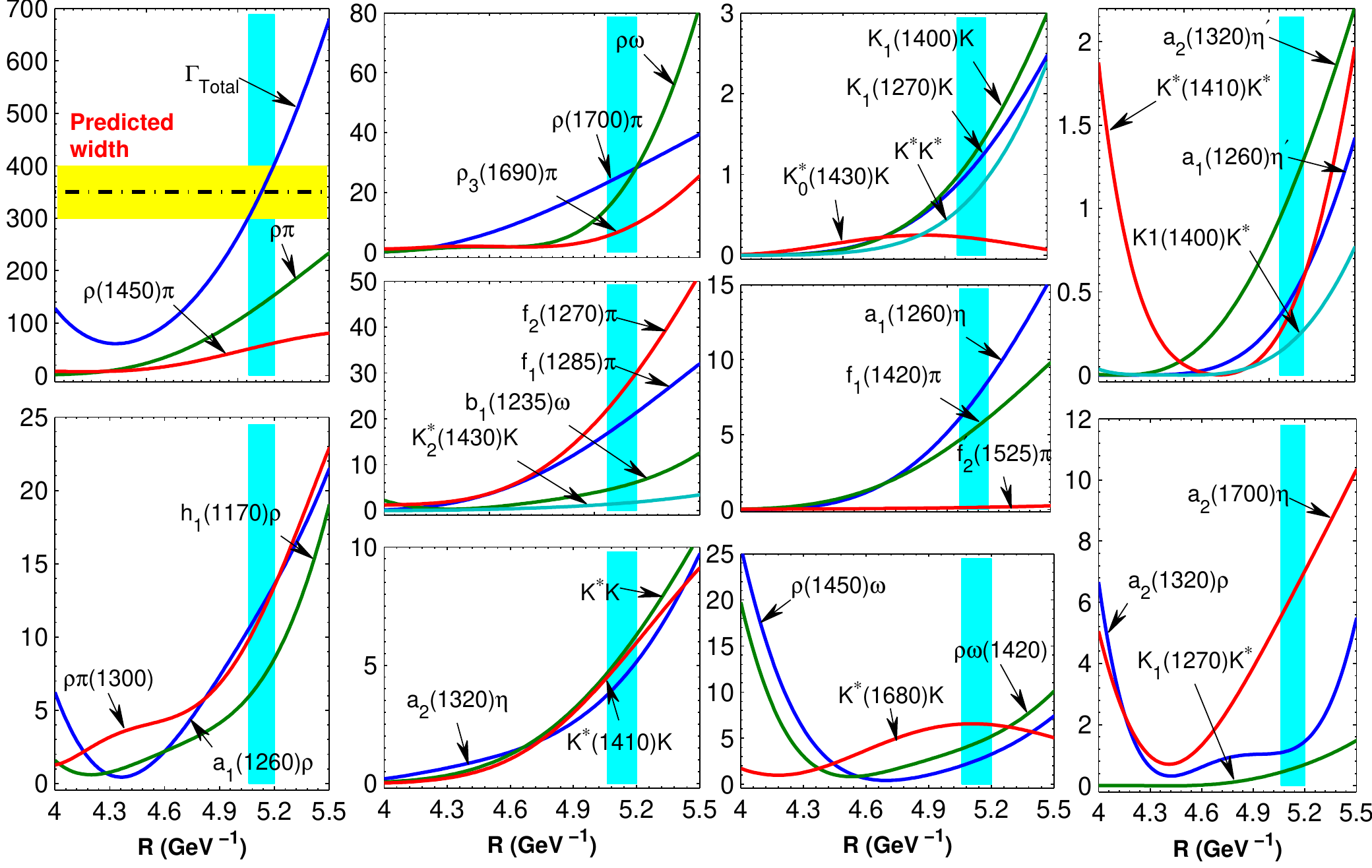}}
\caption{(color online). The $R$ dependence of {the} calculated partial and total decay widths of $\pi_{2}(2540)$.
Here, the dot-dashed line with the horizontal band is the predicted total decay width corresponding to the $R=5.06- 5.2$ GeV$^{-1}$ range. Here, all results are in units of MeV.\label{Pi22540}}
\end{figure*}

As shown in Fig. \ref{Pi22005}, the calculated total decay width of $\pi_2(2005)$ overlaps with the experimental data
in the broad $R$ range due to a large error of the experimental width of $\pi_2(2005)$. Our results further show that $\rho(1450)\pi$, $\rho\pi$, $\rho(1720)\pi$, $\rho\omega$, and $f_2(1270)\pi$ are the main decay modes. However, it depends on the $R$ value whether $\rho_3(1690)\pi$, and $h_1(1170)\rho$ are main decay modes of $\pi_2(2005)$. At present, E852 observed the $\pi_{2}(2005)\to \omega\rho^{-}$ decay \cite{Lu:2004yn}, which can be explained by our calculation.

Figure \ref{Pi22100} gives the information {on} the partial and total decay widths of $\pi_2(2100)$ as the first radial excitation of $\pi_2(1670)$. We find that our results can reproduce the experimental data \cite{Amelin:1995gu}, when taking $R=(4.18- 4.39)$ GeV$^{-1}$, especially the $D$-wave/$S$-wave ratio for $\pi_{2}(2100)\to f_{2}(1270)\pi$. That is, $\pi_{2}(2100)\to f_{2}(1270)\pi$ can occur via $S$-wave and $D$-wave which leads us to study the $D$-wave/$S$-wave ratio for $\pi_{2}(2100)\to f_{2}(1270)\pi$, whose experimental value is given as $0.39\pm0.23$ \cite{Daum:1980ay}. On the other hand, we obtain this ratio to be $0.1- 0.63$ in this work, which covers the above experimental data \cite{Daum:1980ay}.
In addition, $\rho(1450)\pi$, $\rho\pi$,  $a_1(1260)\rho$, $h_1(1170)\rho$, $f_2(1270)\pi$, $\rho\omega$, $\rho_3(1690)\pi$, and $b_1(1235)\omega$
can be the main decay channels, where $\rho\pi$ and $f_2(1270)\pi$ were observed in the experiment \cite{Daum:1980ay}.

It is obvious that the present experimental status of $\pi_2(2005)$ and $\pi_2(2100)$ is not enough to draw a definite conclusion as to which is more suitable for the best candidate as the first radial excitation of $\pi_2(1670)$. More experimental data of $\pi_2(2005)$ and $\pi_2(2100)$ are needed.

\subsubsection{$\pi_{2}(2285)$ and the predicted $\pi_2(2540)$}

In the following, we present the decay behavior of $\pi_{2}(2285)$ in Fig. \ref{Pi22285}, where $\pi_{2}(2285)$ is the $3^{1}D_2$ state. Since the experimental information {on} $\pi_{2}(2285)$ is scarce, we only compare the obtained total width with the experimental data \cite{Anisovich:2010nh}. We notice that the experimental data can be reproduced when taking $R=(4.3- 4.67)$ GeV$^{-1}$. The corresponding main decay channels include $\rho(1450)\pi$, $\rho\pi$, $f_2(1950)\pi$, and $\rho\omega(1420)$. In Table \ref{Pi22285BrTable}, we list some typical ratios of partial decay widths, which can be useful for further experimental study of $\pi_{2}(2285)$

\begin{table}[htb]
\caption{The typical ratios relevant to the decay behavior of $\pi_{2}(2285)$ as the $3^{1}D_{2}$ state and the predicated $\pi_2(2540)$ as the $4^{1}D_{2}$ state with the $R$ ranges  ($4.30-4.67$) GeV$^{-1}$ and ($5.06-5.20$) GeV$^{-1}$, respectively. Here, $\Gamma_{Total}$ denotes the total decay width. \label{Pi22285BrTable}}
\begin{center}
 \renewcommand{\arraystretch}{1.5}
 \tabcolsep=1.4pt
\begin{tabular}{lcc}
\toprule[1pt]
\toprule[1pt]
Ratios&                                  $\pi_{2}(2285)$&                              $\pi_2(2540)$\\
\midrule[1pt]

$\rho\pi/\rho(1450)\pi$&                            $1.0-1.6$&                       $2.3-2.5$\\

$f_{2}(1270)\pi/\rho\pi$&                           $0.14-0.25$&                     $0.18-0.2$\\

$f_{1}(1285)\pi/\rho(1700)\pi$&                     $0.46-0.53$&                     $0.72-0.76$\\

$[\bar{K^{\ast}}K+c.c]/f_{2}(1270)\pi$&             $0.21-0.35$&                      $0.20-0.21$\\

$\rho\omega/\Gamma_{Total}$&                        $0.023-0.029$&                    $0.04-0.07$\\

$\rho(1450)\pi/\Gamma_{Total}$&                     $0.10-0.15$&                      $0.14-0.17$\\

$b_{1}(1235)\pi/\Gamma_{Total}$&                    0&                                   0\\

$a_{2}(1320)\eta/f_{1}(1285)\pi$&                   $0.42-0.67$&                      $0.22-0.24$\\

$\rho\omega/\rho\pi$&                               $0.09-0.27$&                      $0.12-0.18$\\

$\rho(1700)\pi/\Gamma_{Total}$&                     $0.026-0.058$&                      $\sim0.07$\\

$b_{1}(1235)\omega/f_{2}(1270)\pi$&                 $0.21-0.58$&                      $\sim0.2$\\

$a_{2}(1320)\eta/\Gamma_{Total}$&                   $0.008-0.013$&                    $\sim0.12$\\

\bottomrule[1pt]
\bottomrule[1pt]
\end{tabular}
\end{center}
\end{table}

In addition, we also study the OZI-allowed two-body decays of the predicted $\pi_2(2540)$. Since $\pi_2(2540)$ is the higher radial excited state, the partial and total decay widths are strongly dependent on the $R$ value, and hence it is difficult to conclude whether $\pi_2(2540)$ is a broad state or not. Usually the $R$ value becomes larger with increasing the radial quantum number. Thus, if taking a typical $R$ range (see the vertical band in Fig. \ref{Pi22540}), we estimate that  $\pi_2(2540)$ is a broad state with a width around 350 MeV. The corresponding dominant decay modes are $\rho \pi$ and $\rho(1450)\pi$, where the detailed decay information can be found in Fig. \ref{Pi22540} and Table \ref{Pi22285BrTable}, which will be helpful to further experimentally search for this predicted $\pi_2$ state.

\subsection{$\eta_{2}$ meson family}

There are four $\eta_2$ states listed in PDG \cite{Beringer:1900zz}, which are isoscalar. In the following, we mainly introduce their experimental status.

The Crystal Barrel Collaboration studied the $p\bar{p}\to \eta\pi^{0}\pi^{0}\pi^{0}$ reaction \cite{Adomeit:1996nr} and observed two $2^{-+}$ states, $\eta_{2}(1645)$ and $\eta_{2}(1870)$, in the $\eta\pi\pi$ channel, where $\eta_{2}(1645)$ as a partner of $\pi_{2}(1670)$ has the mass $M=1645\pm14\pm15$ and width $\Gamma=180^{+40}_{-21}\pm25$.  $\pi_{2}(1670)$ decays dominantly into $a_{2}(1320)\pi$ via $S$-wave. Later, the WA102 Collaboration confirmed $\eta_{2}(1645)$ in the $a_{2}(1320)\pi$ channel \cite{Barberis:1997ve}, i.e.,
in the reaction $p\bar{p}\to p_{f}(\pi^{+}\pi^{-}\pi^{+}\pi^{-})p_{s}$, a $J^{P}=2^{-}$ signal around 1.6 GeV was observed in the $a_{2}(1320)\pi$ channel, which is consistent with the former result given by the Crystal Barrel Collaboration  \cite{Adomeit:1996nr}. Until now, the observed decay channels of $\eta_{2}(1645)$ are $a_2(1320)\pi$, $K\bar{K}\pi$, $K^*\bar{K}$, $\eta\pi^+\pi^-$, and $a_0(980)\pi$ \cite{Beringer:1900zz}.

In the $\gamma\gamma$ scattering reaction, a $J^{PC}=2^{-+}$ state with mass at 1.9 GeV was announced by the Crystal Barrel Collaboration \cite{Karch:1991sm,Karch:1990xs}, where this state can be described by resonance parameters $M=1881$ and $\Gamma=221$ MeV. In 1996, the evidence for two isoscalar $J^{PC}=2^{-+}$ states at 1645 and 1875 MeV was revealed by the Crystal Barrel Collaboration \cite{Adomeit:1996nr} mentioned above. As for the second signal, it is just above the threshold of $f_{2}(1270)\eta$, and can be well fitted with the mass $1875\pm20\pm35$ MeV and width $\Gamma=250\pm25\pm45$ MeV \cite{Adomeit:1996nr}.
Subsequently, in the decay channels, $a_{0}(980)\pi$, $a_{2}(1320)\pi$ and $f_{2}(1270)\eta$, the WA102 Collaboration confirmed the existence of $\eta_{2}(1870)$ \cite{Barberis:1999be}.
In 2011, Anisovich {\it et al.} reanalyzed the experimental data of the reaction $p\bar{p}\to \eta3\pi^{0}$ collected by the Crystal Barrel and WA102 Collaborations, where $\eta_{2}(1870)$ was reconfirmed \cite{Anisovich:2010nh}.

Although $\eta_2(1870)$ was confirmed by different experiments (see PDG \cite{Beringer:1900zz} for more detailed experimental information), there are difference theoretical explanations for this controversial state. As presented in Fig. \ref{ReggeTraj}, $\eta_2(1870)$ is too light to be the first radial excitation of $\eta_2(1645)$. However, the mass of $\eta_2(1870)$ falls into the predicted mass (around 1.9 GeV) of a $2^{-+}$ hybrid \cite{Isgur:1984bm}, which inspired theorists to explain $\eta_2(1870)$  as the hybrid state \cite{Anisovich:2000mv,Karch:1991sm,Barnes:2002mu,Barnes:1996ff,Amsler:2008zzb,Close:1994hc,Page:1998gz}. Additionally, no evidence of a decay mode of $\eta_{2}(1870)\to {K^\ast}\bar K$ shows that possibility of $\eta_{2}(1870)$ being the $s\bar{s}$ partner of $\eta_{2}(1645)$ and $\pi_{2}(1670)$ can be excluded \cite{Barnes:2002mu}. In Ref. \cite{Li:2009rka},  $\eta_2(1870)$ as the $2^{1}D_{2}$ $n\bar{n}$ state was suggested, however, some important partial decay width was listed in Table I of Ref. \cite{Li:2009rka}. From this table we find the theoretical branching ratio of $K^{\ast}\bar{K}/f_{2}(1270)\eta\approx 1$, which also contradicts with the present experimental fact of the absence of the $K^{\ast}\bar K$ decay mode for $\eta_{2}(1870)$.

In the $p\bar{p}$ annihilation, two $2^{-+}$ resonances above 2 GeV were first reported in Ref. \cite{Anisovich:1999jw}. The first one has the mass $M=2040\pm40$ MeV and width $\Gamma=190\pm40$ MeV, which decays strongly into $f_{2}(1270)\eta$ and weakly couples to $a_{2}(1320)\pi$. In the $p\bar{p}\to \eta3\pi^{0}$ reaction, a similar structure was observed and it decays dominantly into $a_{2}(1320)\pi$ via a $D$-wave and slightly into $a_{2}(1320)\pi$ through an $S$-wave \cite{Anisovich:2000mv}. This structure is named $\eta_2(2030)$ in PDG \cite{Beringer:1900zz}.

Besides $\eta_2(2030)$, another structure with the mass $M=2300\pm40$ and width $\Gamma=270\pm40$
was observed in Ref. \cite{Anisovich:1999jw} by analyzing the $p\bar{p}\to \pi^{0}\pi^{0}\eta$ reaction, which corresponds to $\eta_{2}(2250)$, which decays dominantly into $a_{2}(1320)\pi$. Moreover, the data on $p\bar{p}\to \eta^{'}\pi^{0}\pi^{0}$ were studied, which shows a $2^{-+}$ signal existing in the $f_{2}(1270)\eta^{'}$ invariant mass spectrum \cite{Anisovich:2000us}, which has the mass $M=2248\pm20$ MeV and width $\Gamma=280\pm20$ MeV. The decay modes, $f_{2}(1270)\eta$, $a_{2}(1320)\pi$ and $a_{0}(980)\pi$, of $\eta_{2}(2250)$ were observed when reanalyzing the data on $p\bar{p}\to \eta\pi^{0}\pi^{0}\pi^{0}$ \cite{Anisovich:2000ut}. At present, $\eta_{2}(2250)$ is listed in PDG \cite{Beringer:1900zz} as the further state.

In the following subsections, we perform the phenomenological analysis of $\eta_{2}(1645)$, $\eta_2(2030)$, $\eta_2(2250)$, and a predicted $\eta_2(2480)$, where we treat the discussed $\eta_2$ as pure $n\bar{n}$ states.

\subsubsection{$\eta_{2}(1645)$}

The analysis of the Regge trajectories indicates that $\eta_{2}(1645)$ is a ground state in the $\eta_2$ meson family, which can be the partner of $\pi_2(1670)$.

\begin{figure}[htb]
\scalebox{1.}{\includegraphics[width=\columnwidth]{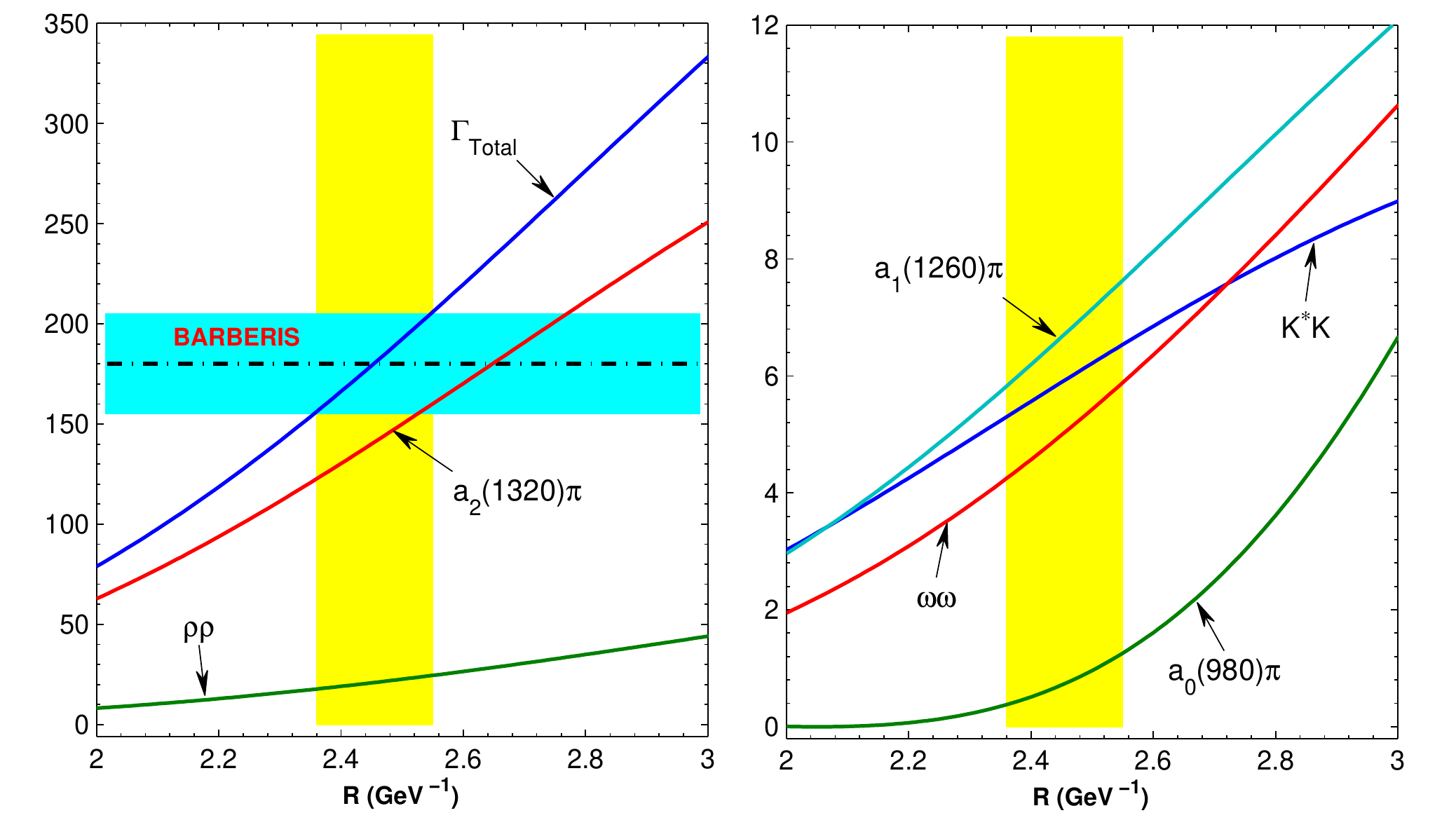}}
\caption{(color online). The $R$ dependence of {the} calculated partial and total decay widths (in units of MeV) of $\eta_{2}(1645)$. Here, the dot-dashed line with the horizontal is the experimental width from Ref.
\cite{Barberis:1997ve}.   The vertical band corresponds to the $R$ range, where the theoretical total decay width overlaps with the experimental data. \label{Eta21645}}
\end{figure}

In Fig. \ref{Eta21645}, the obtained partial and total decay widths of $\eta_2(1645)$ are given by varying $R$ and are compared with the experimental widths \cite{Barberis:1997ve}. When $R=(2.36-2.55)$ GeV$^{-1}$, the calculated total decay width can overlap with the experimental data. Here, we need to mention that the $R$ range for  $\eta_2(1645)$ is similar to that for $\pi_2(1670)$.
Furthermore, the partial decay information indicates that $a_2(1320)\pi$ is a dominant decay channel of  $\eta_{2}(1645)$, which is consistent with the experimental observation \cite{Adomeit:1996nr}.

Another our calculation is the ratio $\Gamma(K^*\bar{K})/\Gamma(a_2(1320)\pi)$, which gives this value is $0.038-0.043$, which is comparable with
the experimental result of $\Gamma(K\bar{K}\pi)/\Gamma(a_2(1320)\pi)=0.07\pm0.02\pm0.02$ \cite{Barberis:1997vf}. This fact shows that the assumption of $\eta_2$ as pure $n\bar{n}$ states is reasonable. We also calculate
the $D$-wave/$S$-wave ratio for $\eta_{2}(1645)\to a_{2}(1320)\pi$, which is about 0.016$\sim$0.018 and is consistent with the experimental results since $\eta_{2}(1645)$ decays dominantly into $a_{2}(1320)\pi$ via the $S$-wave \cite{Barberis:1997ve}.

Thus, our study supports $\eta_{2}(1645)$ as the pure $1^{1}D_{2}$ $n\bar{n}$ state.

\subsubsection{$\eta_{2}(2030)$}

Assuming $\eta_{2}(2030)$ is the first radial excitation of $\eta_{2}(1645)$ (see the analysis shown in Fig. \ref{ReggeTraj}),
we study the partial and total decay widths, which are illustrated in Fig. \ref{Eta22030}.

\begin{figure}[htb]
\scalebox{0.97}{\includegraphics[width=\columnwidth]{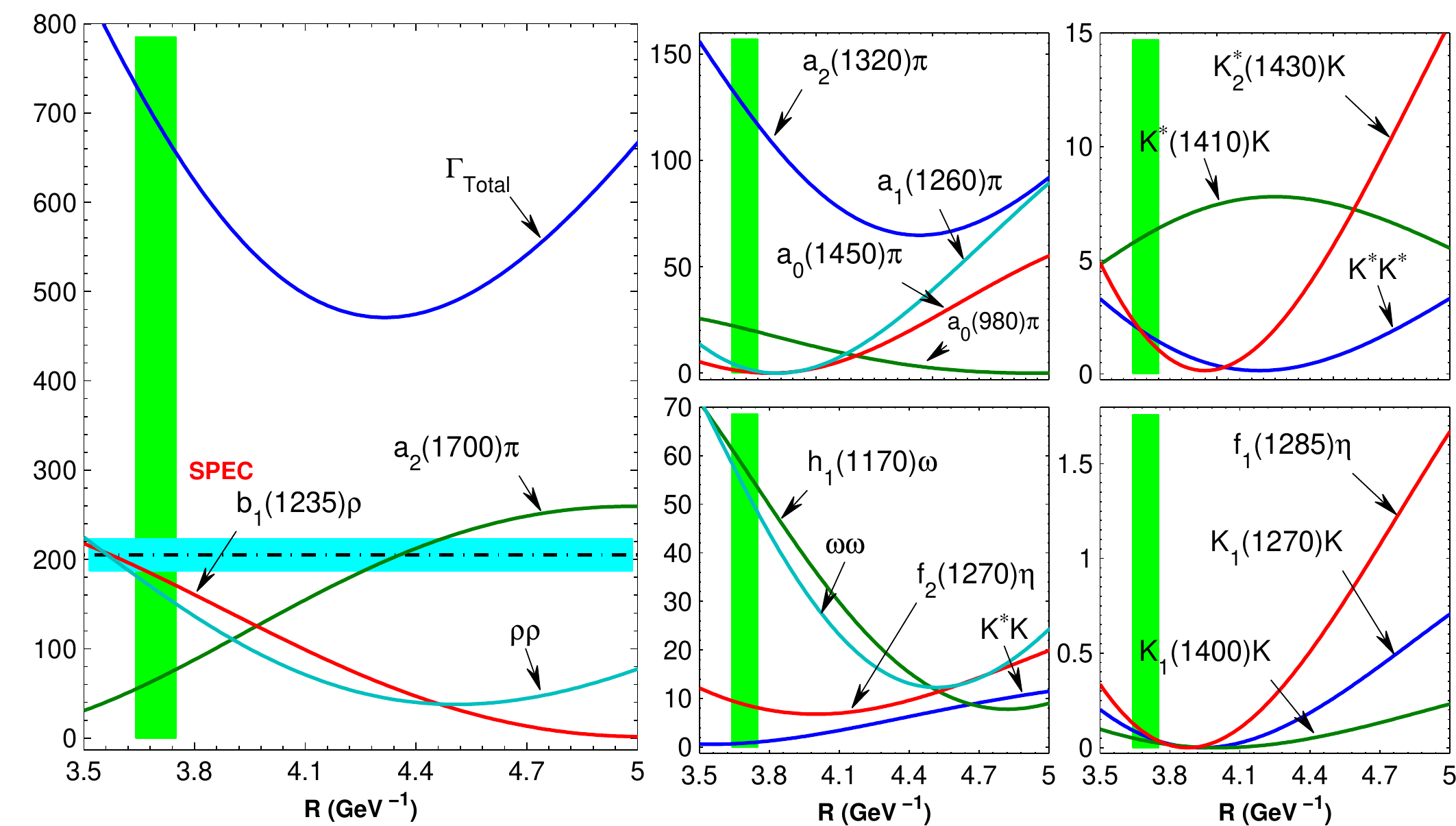}}
\caption{(color online). The $R$ dependence of {the} calculated partial and total decay widths of $\eta_{2}(2030)$. Here, the dot-dashed line with the horizontal band denotes the experimental width from Ref.
\cite{Anisovich:2000mv}.\label{Eta22030}}
\end{figure}

Our results show that the obtained total decay width of $\eta_{2}(2030)$ is far larger than the experimental result given in Ref. \cite{Anisovich:2000mv}. At present, $\eta_{2}(2030)$ is as a further state listed in PDG  \cite{Beringer:1900zz}, and the corresponding experimental information is not enough to clarify this discrepancy. We suggest further experiment to measure the resonance parameters of $\eta_{2}(2030)$, which will reveal the underlying properties of $\eta_{2}(2030)$.

\begin{table}[htbp]
\caption{The obtained typical ratios relevant to the decays of $\eta_{2}(2030)$ as the $2^{1}D_{2}$ state, where we take the typical $R= (3.64-3.75)$ GeV$^{-1}$ range. The experimental data from Ref. \cite{Anisovich:2000mv}. Here, $L=0$ and $L=2$ denote that the corresponding decays occur via $S$-wave and $D$-wave, respectively.\label{Eta22030BrTable}}
\begin{center}
 \renewcommand{\arraystretch}{1.5}
 \tabcolsep=1.4pt
\begin{tabular}{lcc}
\toprule[1pt]
\toprule[1pt]
Ratios&                                  This work&                    Experimental data\\
\midrule[1pt]
$\Gamma(a_{2}(1320)\pi)_{L=0}/\Gamma(a_{2}(1320)\pi)_{L=2}$&                       0.57$-$0.9&               0.74$\pm$0.17\\

$\Gamma(a_{0}(980)\pi)/\Gamma(a_{2}(1320)\pi)_{L=2}$&                             0.33$-$0.41&                0.37$\pm$0.08\\

$\Gamma(f_{2}(1270)\eta)/\Gamma(a_{2}(1320)\pi)_{L=2}$&                           0.14$-$0.17&                 0.15$-$0.43\\

\bottomrule[1pt]
\bottomrule[1pt]
\end{tabular}
\end{center}
\end{table}

In Table \ref{Eta22030BrTable}, we list three typical ratios and the comparison with the experimental results in Ref. \cite{Anisovich:2000mv}, which shows that the experimental data can be well reproduced by our calculations.
The results in Fig. \ref{Eta22030} provide the information {on the} main decay modes of $\eta_{2}(2030)$. If we take $R= (3.64-3.75)$ GeV$^{-1}$ as a typical range to discuss this point, we find that $b_1(1235)\rho$, $\rho\rho$, $a_2(1700)\pi$, $a_2(1320)\pi$, $h_1(1170)\omega$, and $\omega\omega$ are its main decay channels, which are valuable for further experimental study on $\eta_{2}(2030)$.


\subsubsection{$\eta_{2}(2250)$ and the predicted $\eta_{2}(2480)$}

Under the assignment of the $3^{1}D_{2}$ state to $\eta_{2}(2250)$, we discuss the decay behavior of $\eta_{2}(2250)$, which is presented in Fig. \ref{Eta22250}. Our theoretical result can well reproduce the experimental width of  $\eta_{2}(2250)$ \cite{Anisovich:2000ut} when taking $R=(4.95\sim5.17)$ GeV$^{-1}$, which is comparable with the former obtained $R$ range for $\pi_2(2285)$. Furthermore, the main decay channels of $\eta_{2}(2250)$ were obtained, i.e., $a_2(1700)\pi$, $a_1(1260)\pi$, $\rho\rho$, $a_2(1320)\pi$, $\omega\omega$ and $b_1(1235)\rho$. Comparing with the former discussed three $\eta_2$ states, the experimental information is insufficient since experiment measured only the resonance parameters. More experimental study of $\eta_{2}(2250)$ is helpful to establish this $\eta_2$ state listed as a further state in PDG \cite{Beringer:1900zz}. In Table \ref{Eta22250BrTable}, we further provide some predicted ratios relevant to the partial decay widths, which can be tested in future experiment.

\begin{figure}[htb]
\scalebox{1.}{\includegraphics[width=\columnwidth]{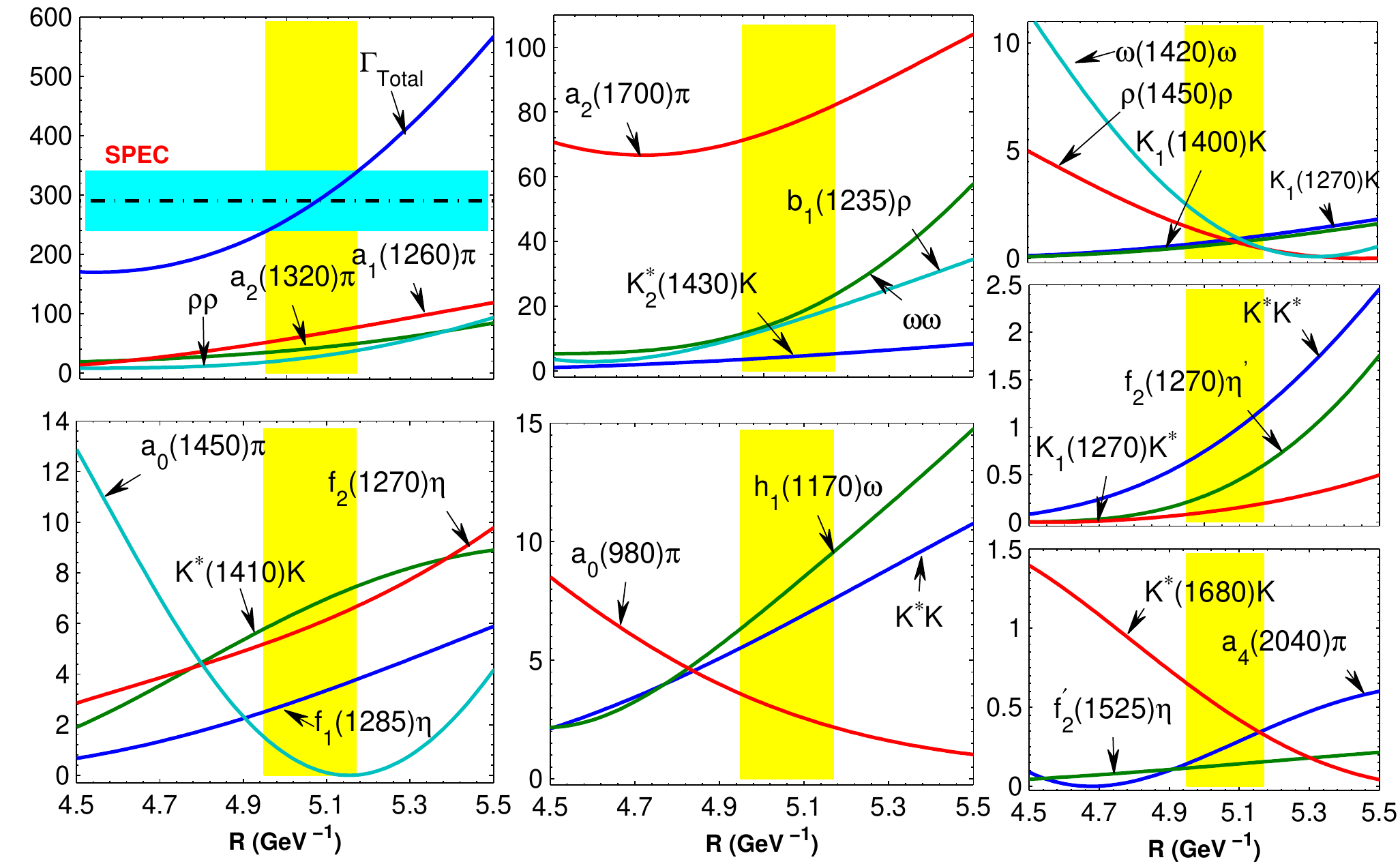}}
\caption{(color online). The variation of {the}  calculated partial and total decay widths of $\eta_{2}(2250)$
in the $R$ value. Here, the
dot-dashed line with the horizontal band is the experimental total width from Ref.
\cite{Anisovich:2000ut}. All decay widths are in units of MeV. The vertical band means that the range of the theoretical result overlaps with the experimental data.\label{Eta22250}}
\end{figure}

\begin{table}[htb]
\caption{The typical ratios relevant to the decay behavior of $\eta_{2}(2250)$ as the $3^{1}D_{2}$ state and the predicted $\eta_2(2480)$ as $4^{1}D_{2}$ state with the $R$ ranges (4.95$-$5.17) GeV$^{-1}$ and (5.36$-$5.49) GeV$^{-1}$, respectively. Here, $\Gamma_{Total}$ denotes the total decay width.  \label{Eta22250BrTable}}
\begin{center}
 \renewcommand{\arraystretch}{1.5}
 \tabcolsep=1.4pt
\begin{tabular}{lcc}
\toprule[1pt]
\toprule[1pt]
Ratios&                                      $\eta_{2}(2250)$ &                                  $\eta_{2}(2480)$\\
\midrule[1pt]

$a_{1}(1260)\pi/\Gamma_{Total}$&                        0.21$-$0.23&                        0.23$-$0.25\\

$a_{2}(1320)\pi/\Gamma_{Total}$&                       0.14$-$0.15&                       0.18$-$0.19\\

$[\bar{K^{\ast}}(892)K+c.c]/f_{2}(1270)\eta$&           $\approx$ 1&                           1.0$-$1.1\\

$\rho\rho/a_{2}(1320)\pi$&                             0.52$-$0.76&                         0.7$-$0.9\\

$a_{0}(980)\pi/f_{2}(1270)\eta$&                       0.33$-$0.70&                         0.24$-$0.33\\

$a_{2}(1320)\pi/f_{1}(1285)\eta$&                      13.2$-$13.8&                         10.3$-$10.8\\

$\omega\omega/f_{2}(1270)\eta$&                        2.2$-$3.5&                           2.4$-$2.9\\

$b_{1}(1235)\rho/\Gamma_{Total}$&                      0.044$-$0.057&                       $\approx$ 0.56\\

$a_{2}(1700)\pi/\Gamma_{Total}$&                       0.24$-$0.30&                        0.1$-$0.12\\

$K_1(1270)K/a_{2}(1320)\pi$&                           0.018$-$0.02&                        $\approx$ 0.02\\

$\omega\omega/\Gamma_{Total}$&                         0.05$-$0.07&                         0.04$-$0.05\\

\bottomrule[1pt]
\bottomrule[1pt]
\end{tabular}
\end{center}
\end{table}

In addition, the decay behavior of the predicted $\eta_2(2480)$ state with a $4^1D_2$ quantum number
is { crucial information for future experimental searches} for this predicted state, which are listed in Fig. \ref{Eta22480} and Table \ref{Eta22250BrTable}. The calculated total decay width of $\eta_{2}(2480)$ is sensitive to
the $R$ value, which is mainly due to a node effect, where the situation of $\eta_{2}(2480)$ is similar to that of $\pi_2(2540)$. To quantitatively discuss the decay behavior of the predicted $\eta_{2}(2480)$, we take $R=(5.36-5.49)$ GeV$^{-1}$ as the typical range since the $R$ value becomes larger when the radial quantum number is increased. Under this situation, we predict that $\eta_{2}(2480)$ is a broad state with the width around 400 MeV, where the main decay modes include $a_1(1260)\pi$, $\rho\rho$, $a_2(1320)\pi$ and $a_2(1700)\pi$.

\begin{figure*}[htbp]
\scalebox{1.5}{\includegraphics[width=\columnwidth]{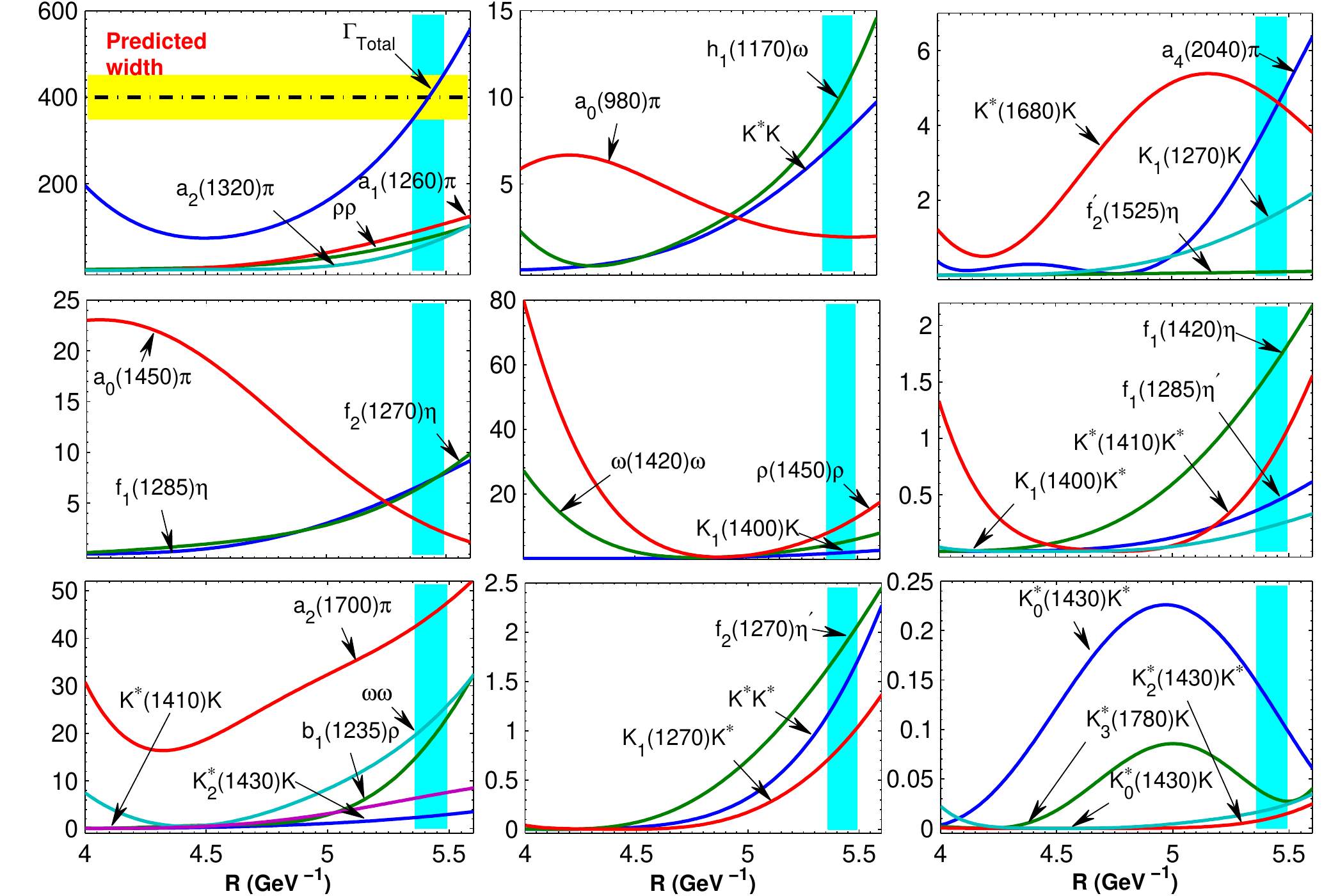}}
\caption{(color online). The $R$ dependence of {the} calculated partial and total decay widths of $\eta_{2}(2480)$ (in units of MeV). \label{Eta22480}}
\end{figure*}

\subsubsection{Possibility of $\eta_{2}(1870)$ as a partner of $\eta_2(1645)$}
{{
As indicated by the analysis of the $(J, M^2)$ plots in Fig. \ref{EtaJM2}, there exists a possibility that $\eta_2(1870)$
is a partner of $\eta_2(1645)$, which satisfies the following relation
\begin{eqnarray}
\label{FlavorMixingEquation}
\left(\begin{array}{c}|\eta_2(1645)\rangle \\|\eta_2(1870)\rangle\end{array}\right)&=&\left(\begin{array}{cc}\cos\theta&-\sin\theta \\ \sin\theta&\cos\theta\end{array}\right)\left(\begin{array}{c}|n\bar{n}\rangle \\|s\bar{s}\rangle\end{array}\right),
\end{eqnarray}
where $\theta$ is the mixing angle.

After assigning $\eta_{2}(1870)$ to be a partner of $\eta_2(1645)$, we study the decay behavior of $\eta_{2}(1870)$, which is shown in Fig. \ref{eta21870mixing}, where all the results depend on the mixing angle $\theta$. We compare the calculated total decay width with the experimental data \cite{Barberis:1999be}, where the theoretical result overlaps with the experimental data when taking the small mixing angle $\theta$, which shows $\eta_{2}(1870)$ is dominated by the $s\bar{s}$ component. Our results of the partial decay widths also indicate that the $K^*\bar{K}$ mode is the main decay channel of $\eta_{2}(1870)$. At present, it is a puzzling feature that $\eta_{2}(1870)\to {K^\ast}\bar K$ is still missing in experiment \cite{Beringer:1900zz}, which is waiting for the solution from a future experimental and theoretical joint effort.

\begin{figure}[htbp]
\scalebox{0.97}{\includegraphics[width=\columnwidth]{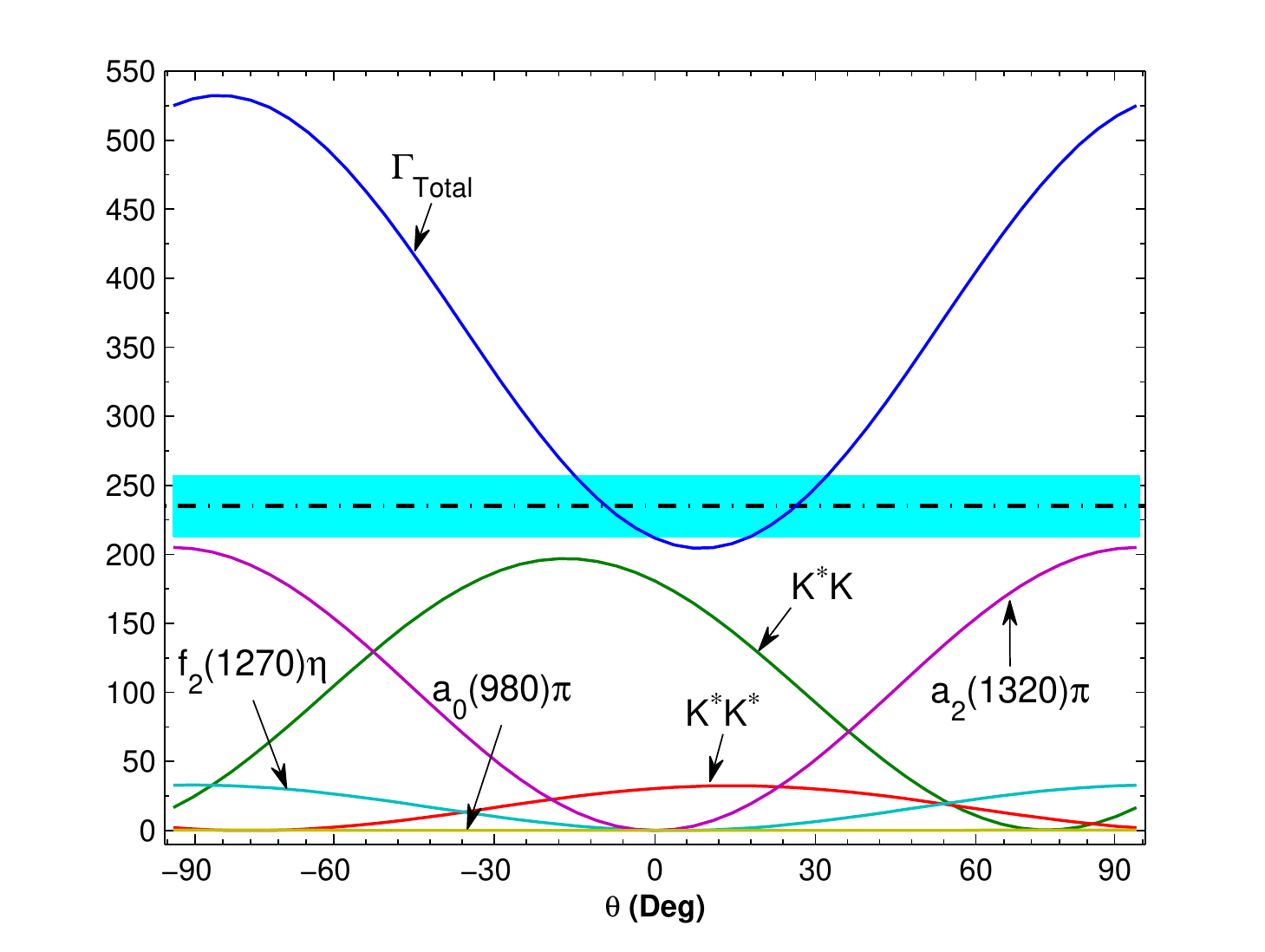}}
\caption{(color online). {The $\theta$ dependence of {the} calculated partial and total decay widths of $\eta_{2}(1870)$. The dot-dashed line with the horizontal band denotes the experimental width from Ref.
\cite{Barberis:1999be}. Here, we fix $R=2.5$ GeV$^{-1}$ for $\eta_{2}(1870)$. \label{eta21870mixing}}}
\end{figure}
}}

\subsection{$K_{2}$ meson family}

As listed in PDG \cite{Beringer:1900zz}, four $K_2$ states with spin-parity $J^P=2^-$ are collected, which are $K_2(1580)$, $K_2(1770)$, $K_2(1820)$, and $K_2(2250)$. Before carrying out the study of these $K_2$ states, we briefly review their experimental status.

In 1966, the evidence for $K_{2}(1770)$ was first reported in the reaction $K^{-}p\to pK^{-}\pi^{+}\pi^{-}$ and $K^{-}p\to p\bar{K^{0}}\pi^{-}\pi^{0}$ \cite{Bartsch:1966zz}, where $K_{2}(1770)$ appears in the $K\pi\pi$ invariant mass distribution with the mass $M=1789\pm10$ MeV and width $\Gamma=80^{+20}_{-40}$ MeV. In 1969, a similar enhancement at 1780 MeV was observed in the $K_{2}^{\ast}(1430)\pi$ channel \cite{BarbaroGaltieri:1969sg}. Subsequently, $K_{2}(1770)$ was also reported in its $K_{2}^{\ast}(1430)\pi$ mode by studying the $K\pi\pi$ system in $K^{-}p\to K^{-}\pi^{+}\pi^{-}p$ \cite{Ludlam:1970gq}. In 1981, at least one $I=\frac{1}{2}$, $J^{P}=2^{-}$ meson was established in the diffractive process $K^{-}p\to K^{-}\pi^{-}\pi^{+}p$ \cite{Daum:1981hb}, which couples strongly to $K_{2}^{\ast}(1430)\pi$, $f_{2}(1270)K$ and $K^{\ast}\pi$. Here another $K_2(1820)$ state was reported, which will be introduced later.

In 1993, the evidence for two $J^{P}=2^{-}$ strange states was announced in the reaction $K^{-}p\to K^{-}\pi^{+}\pi^{-}\pi^{0}p$ \cite{Aston:1993qc}, where one state is around 1.77 GeV and another one is located at 1.82 GeV, both of which couple to $K^{-}\omega$ and then $\omega$ decays into $\pi^{+}\pi^{-}\pi^{0}$. The study of $K_{S}K_{S}K_{L}$ system in the collision of $\pi^{-}C\to K_{S}K_{S}K_{L}+Y$ was presented in Ref. \cite{Tikhomirov:2003gg}, where they observed four $2^{-}$ strange states ($K_{2}(1770)$, $K_{2}(1820)$, $K_{2}(1980)$, and $K_{2}(2280)$). The $K_{2}(1770)$ and $K_{2}(1980)$ were observed in the $f_{0}(980)K$ and $f_{2}(1270)K$ modes. However, we should mention that the $K_{2}(1980)$ was not collected into PDG \cite{Beringer:1900zz}. In the present work, this unconfirmed state $K_{2}(1980)$ associated with another unconfirmed $K_{2}(1580)$ are not considered.

In the following, we introduce $K_{2}(2250)$ first reported in Ref. \cite{Tikhomirov:2003gg}. In 1970, D. Lissauer {\it et al.} analyzed the reaction $K^{+}p\to \bar{Y}NN$, where $\bar{Y}$ denotes $\bar{\Lambda}$ or $\bar{\Sigma}$. They found a $I=\frac{1}{2}$ enhancement with the mass $M=2240\pm20$ MeV and width $\Gamma=80\pm20$ MeV in the $\bar{\Lambda}N$ and ${\Sigma}^{\pm}N$ final states \cite{Lissauer:1970ye}. After {nine} years, an amplitudes analysis of the moments shows the evidence of a $J^{P}=2^{-}$ state at 2.3 GeV in the reaction $K^{+}p\to (\bar{\Lambda}p)p$ and $K^{-}p\to (\bar{p}\Lambda)p$ \cite{Cleland:1980ya}. Later, an analysis of $\bar{p}\Lambda$ system in the process of $K^{-}p\to (\Lambda\bar{p})p$ was performed in Ref. \cite{Baubillier:1980uy}, where a strange state with $J^{P}=2^{-}$ was observed, which has the mass $2235\pm50$ MeV, and width $\sim200$ MeV.
In Ref. \cite{Armstrong:1983gt}, the partial wave analysis of experimental data about the previous reaction was carried out by T. Armstrong {\it et al.}, where they reported a structure with spin-parity $2^{-}$, mass $2200\pm40$ MeV, and width $150\pm30$ MeV. By the efforts from the above experiments, $K_{2}(2250)$ was established in experiment and listed in PDG  \cite{Beringer:1900zz}.

\subsubsection{$K_{2}(1770)$ and $K_{2}(1820)$}

The study of the Regge trajectories indicates that both $K_{2}(1770)$ and $K_{2}(1820)$ are the ground state in the $K_2$ meson family, which are mixture of $1^1D_2$ and $1^3D_2$ states as shown in Eq. (\ref{SpinMixingEquation1}), where the mixing angle $\theta_{K(1)}$ is an important input parameter determined by the experimental data. In Ref. \cite{Barnes:2002mu}, Barnes {\it et al.} once adopted the LASS result to fix the mixing angle $\theta_{K(1)}$, i.e., the LASS experiment measured the $F$-wave/$P$-wave amplitude ratio for $K_2(1820)\to \omega K$  \cite{Aston:1993qc}, which is quite small and is related to the mixing angle $\theta_{K(1)}$. By this experimental data, $\theta_{K(1)}$ was determined to be $-39^\circ$ \cite{Barnes:2002mu}. In our following calculation, we take $\theta_{K(1)}=-39^\circ$ to discuss the decay behaviors of $K_{2}(1770)$ and $K_{2}(1820)$.

Figure \ref{K21770} shows that partial and total decay widths of $K_{2}(1770)$ depend on the $R$ value, where $K_2^*(1430)\pi$ is dominant decay channel of $K_{2}(1770)$, which is consistent with the experimental data  \cite{Beringer:1900zz}. Our results also indicate that $K^*\pi$ is the main decay mode and was actually observed in experiment  \cite{Beringer:1900zz}. Comparing the experimental data with theoretical results of the total decay width, we find that we reproduce the experimental width of $K_{2}(1770)$ \cite{Aston:1993qc} when taking $R=2.02$ GeV$^{-1}$, which is a little bit smaller than the $R$ value obtained in studying $\pi_2(1670)$ and $\eta_2(1645)$.


\begin{figure}[htb]
\scalebox{1.0}{\includegraphics[width=\columnwidth]{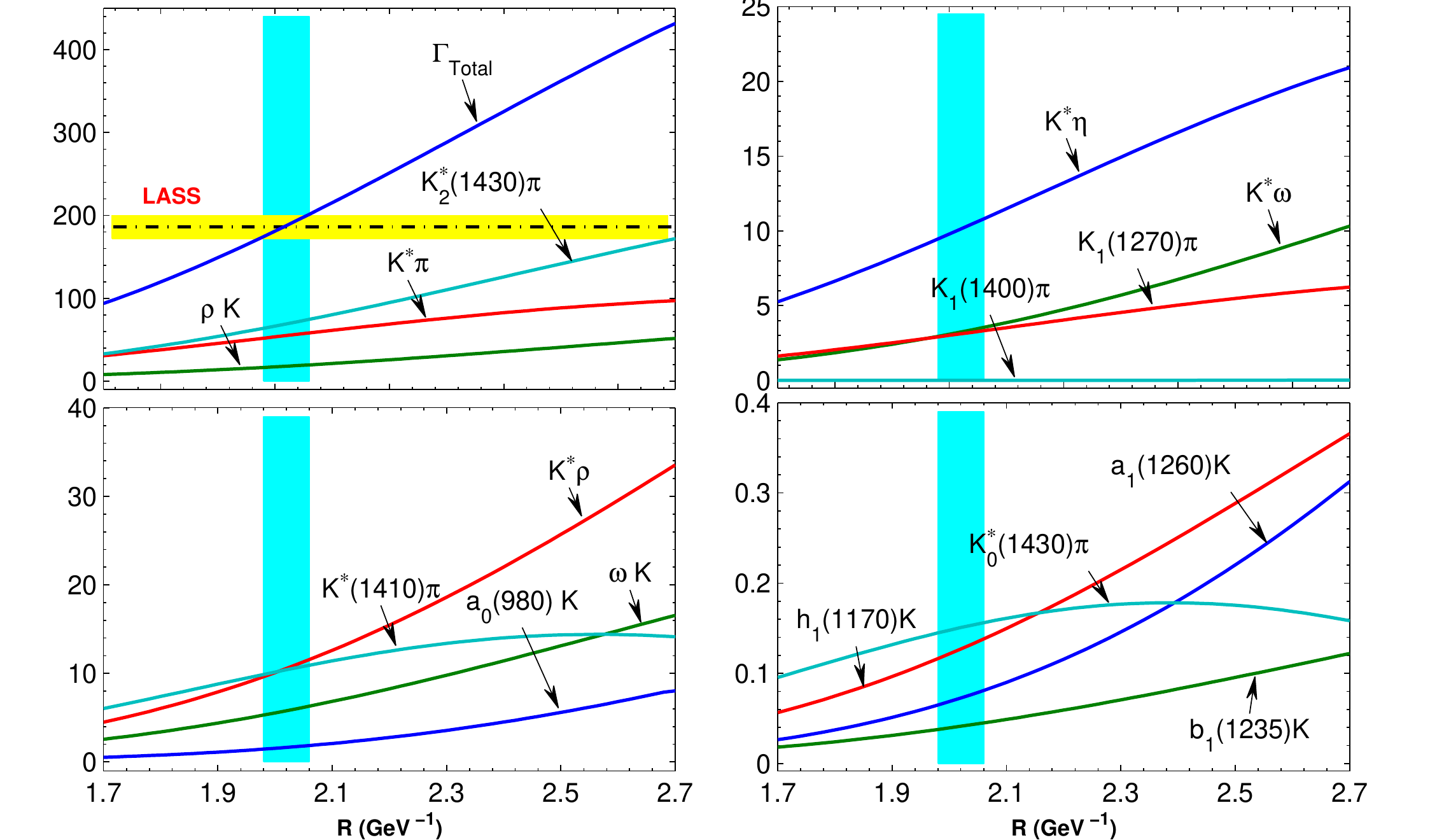}}
\caption{(color online). The $R$ dependence of {the} calculated partial and total decay widths of $K_{2}(1770)$ (in units of MeV). Here, the dot-dashed line with the error band denotes the experimental width of $K_{2}(1770)$ in Ref. \cite{Aston:1993qc}. \label{K21770}}
\end{figure}

\begin{figure}[htb]
\scalebox{1.0}{\includegraphics[width=\columnwidth]{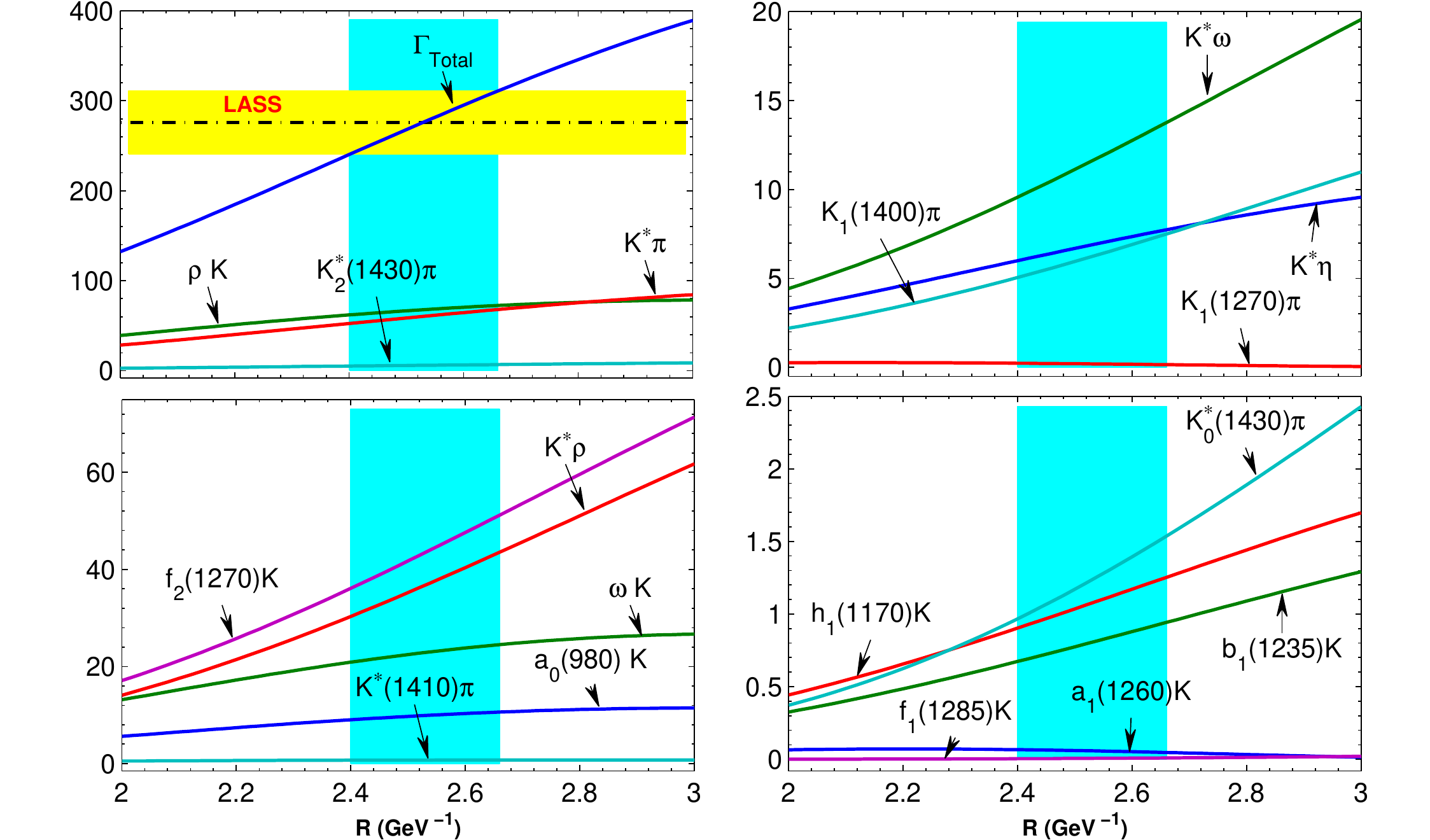}}
\caption{(color online). The $R$ dependence of {the} calculated partial and total decay widths of $K_{2}(1820)$ (in units of MeV). Here, the dot-dashed line with the error band is the experimental width of $K_{2}(1820)$ in Ref. \cite{Aston:1993qc}. \label{K21820}}
\end{figure}

As for $K_{2}(1820)$, the obtained decay behavior is given in Fig. \ref{K21820}, where the calculated total decay width
corresponding to $R=2.54$ GeV$^{-1}$ can describe the experimental data in Ref. \cite{Aston:1993qc}. Here, we need to emphasize that the adopted $R$ range for $K_{2}(1820)$ is comparable with that obtained in investigating $\pi_2(1670)$ and $\eta_2(1645)$, which reflects the requirement that $K_{2}(1820)$, $\pi_2(1670)$ and $\eta_2(1645)$ belong to the same nonet. Additionally, the results in Fig. \ref{K21820} also provide the information of main decay modes, which are $\rho K$, $K^*\pi$, $f_2(1270)K$ and $K^*\rho$. If checking the PDG data \cite{Beringer:1900zz}, we notice that $K_2^*(1430)\pi$, $K^*\pi$, $f_2(1270)K$, and $K\omega$ were observed in experiment, which are also realized by the results listed in Fig. \ref{K21820}.

Due to the above study, we can conclude that $K_{2}(1770)$ and $K_{2}(1820)$ are the ground sates in the $K_2$ meson family. The results shown in Figs. \ref{K21770} and \ref{K21820} provide abundant information of the decays behaviors of $K_{2}(1770)$ and $K_{2}(1820)$, which is useful for future experimental exploration of $K_{2}(1770)$ and $K_{2}(1820)$.

\subsubsection{$K_{2}(2250)$ and the predicted partner $K_{2}(2200)$}

Under the assignment of $K_{2}(2200)/K_{2}(2250)$ as the first radial excitation of $K_{2}(1770)/K_{2}(1820)$,
we illustrate the OZI-allowed two-body decays of $K_{2}(2200)$ and $K_{2}(2250)$, where $K_{2}(2200)$ is a predicted state as the partner of $K_{2}(2250)$, both of which satisfy the relation shown in Eq. (\ref{SpinMixingEquation2}).

In Fig. \ref{K2Figure1}, we present the total decay widths depending on the $R$ value and the mixing angle $\theta_{K(2)}$, where the $R$ range is taken as $R=(3.5- 4.5)$ GeV$^{-1}$, which is from the experience of studying $\pi_2(2005)/\pi_2(2100)$ and $\eta_2(2030)$ since $\pi_2(2005)/\pi_2(2100)$ and $\eta_2(2030)$ with $K_{2}(2250)$ and the predicted $K_{2}(2200)$ form a nonet.

As for $K_{2}(2250)$, the obtained total decay width is far larger than the average experimental width listed in PDG. This discrepancy should be clarified by further precise experimental measurement of the resonance parameters of $K_{2}(2250)$. When taking $R=4.0$ GeV$^{-1}$ and assuming $\theta_{K(2)}=\theta_{K(1)}=-39^\circ$, we obtain the corresponding partial decay widths, which are listed in Table \ref{K2Table1}. The results in Table  \ref{K2Table1} show that $K_2(2250)$ dominantly decays into $K^{\ast}(1410)\pi$ and $K^{\ast}\pi$. In addition, there are several main decay channels, which include $\rho(1450)K$, $b_1(1235)K^*$,  $K_1(1270)\rho$, and $K^*(1410)\rho$. At present, $K\pi\pi$, $K f_2(1270)$, and $K^\ast f_0(980)$ were seen in experiment. We expect that this decay information in Table  \ref{K2Table1}  can be tested in future experiment.

As for the predicted $K_2(2200)$, the two-body strong decay behavior collected in Table \ref{K2Table1} indicates that the $K_1(1270)\rho$, $\rho K$,  $K^{\ast}(1410)\pi$, $K_2^{\ast}(1430)\pi$, and $\rho(1450)K$ modes are {the} main contributions to the total decay width. Thus, we suggest an experiment to search for the predicted $K_2(2200)$ {with} its $\rho K$, $\omega K$, and $K_1(1270)\rho$ decay modes. In addition, when we take the range of {the} $R$ value is $(3.8-4.1)$ GeV$^{-1}$, its typical total decay width can reach up to $486-515$ MeV, which means that the predicted $K_2(2200)$ is a broad state and it is not easy to identify $K_2(2200)$ in experiment.

\begin{figure*}[htbp]
\begin{minipage}{0.45\linewidth}
  \centerline{\includegraphics[width=9cm]{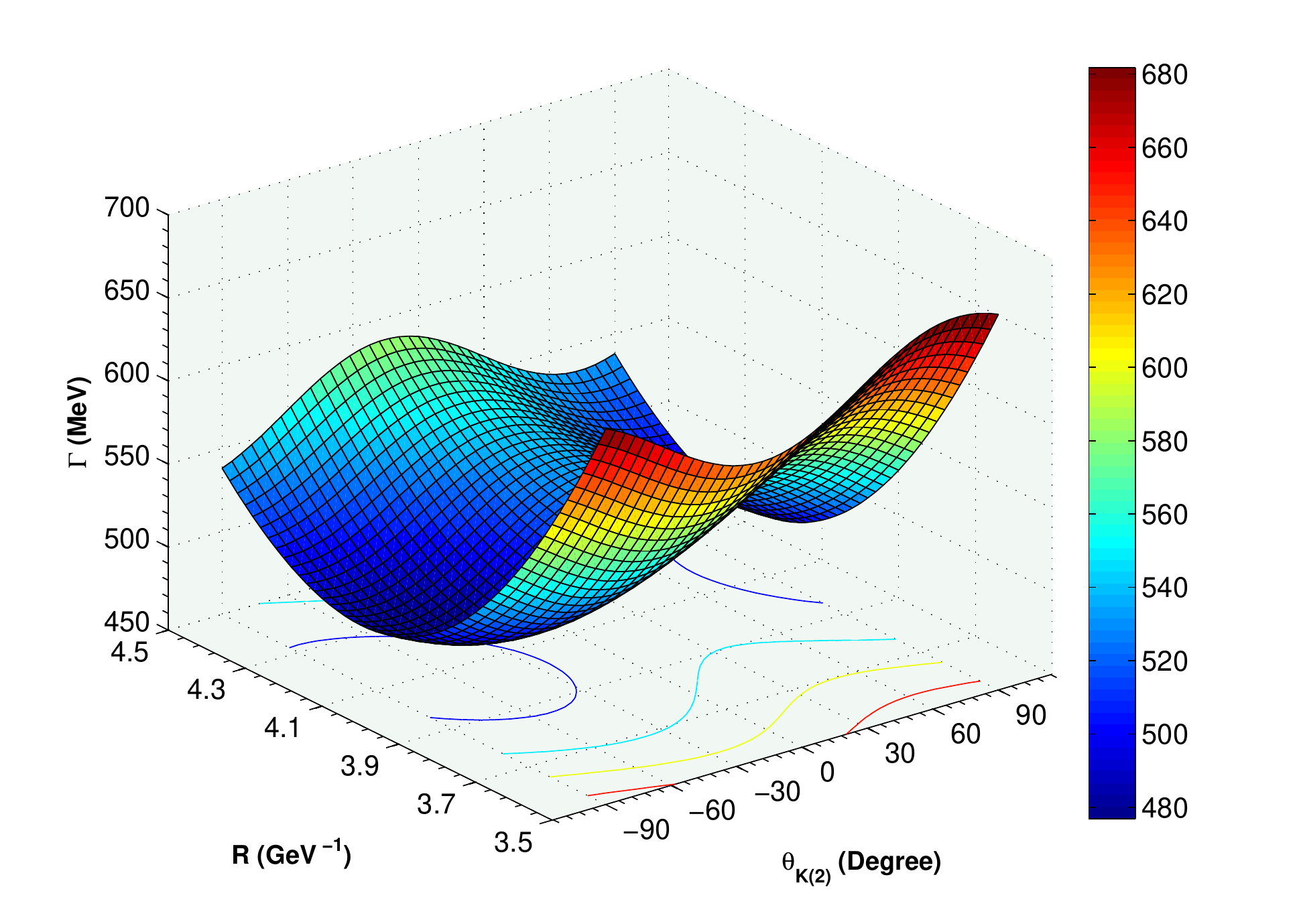}}
\end{minipage}
\hfill
\begin{minipage}{0.45\linewidth}
  \centerline{\includegraphics[width=9cm]{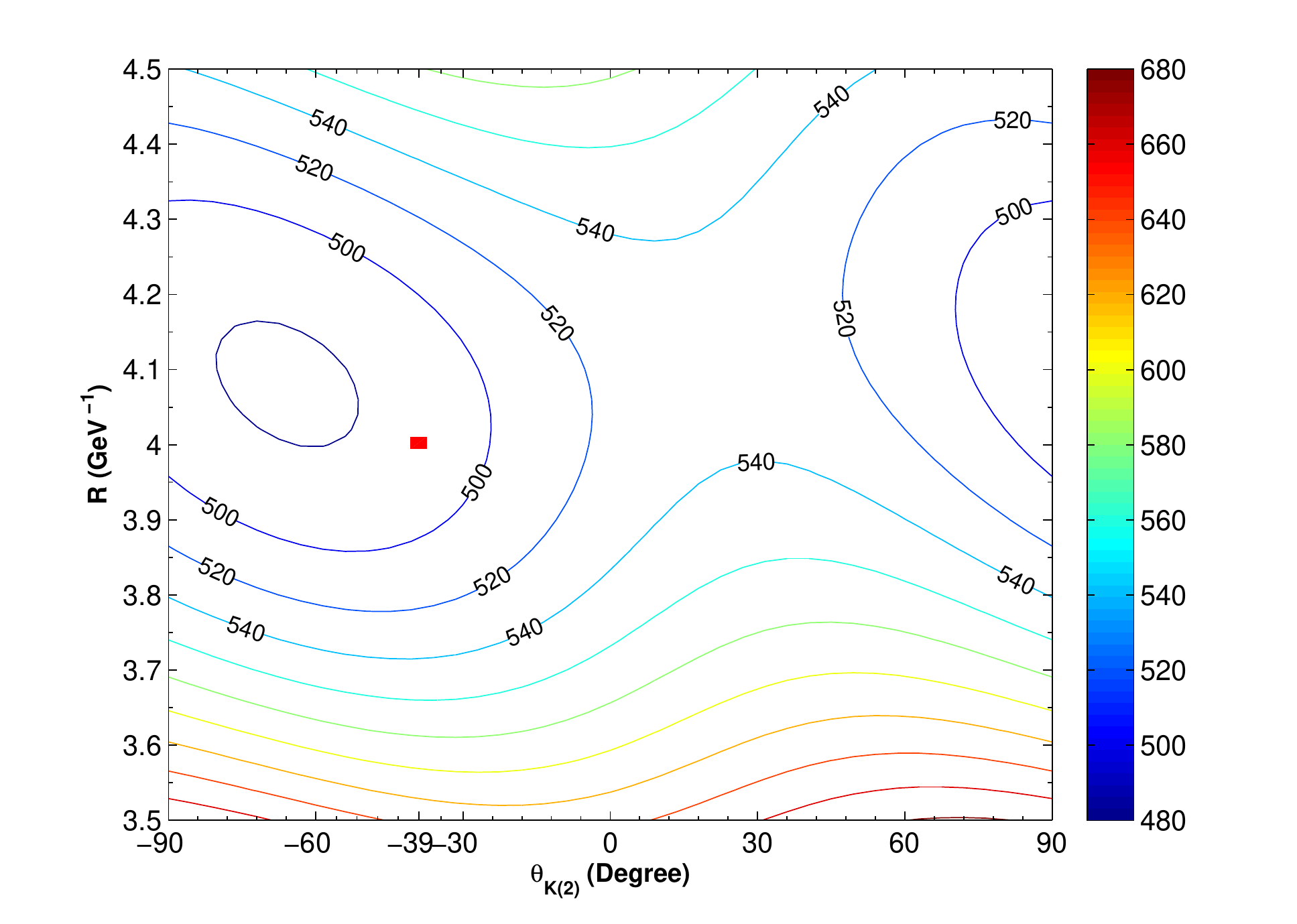}}
\end{minipage}
\vfill
\begin{minipage}{0.45\linewidth}
  \centerline{\includegraphics[width=9cm]{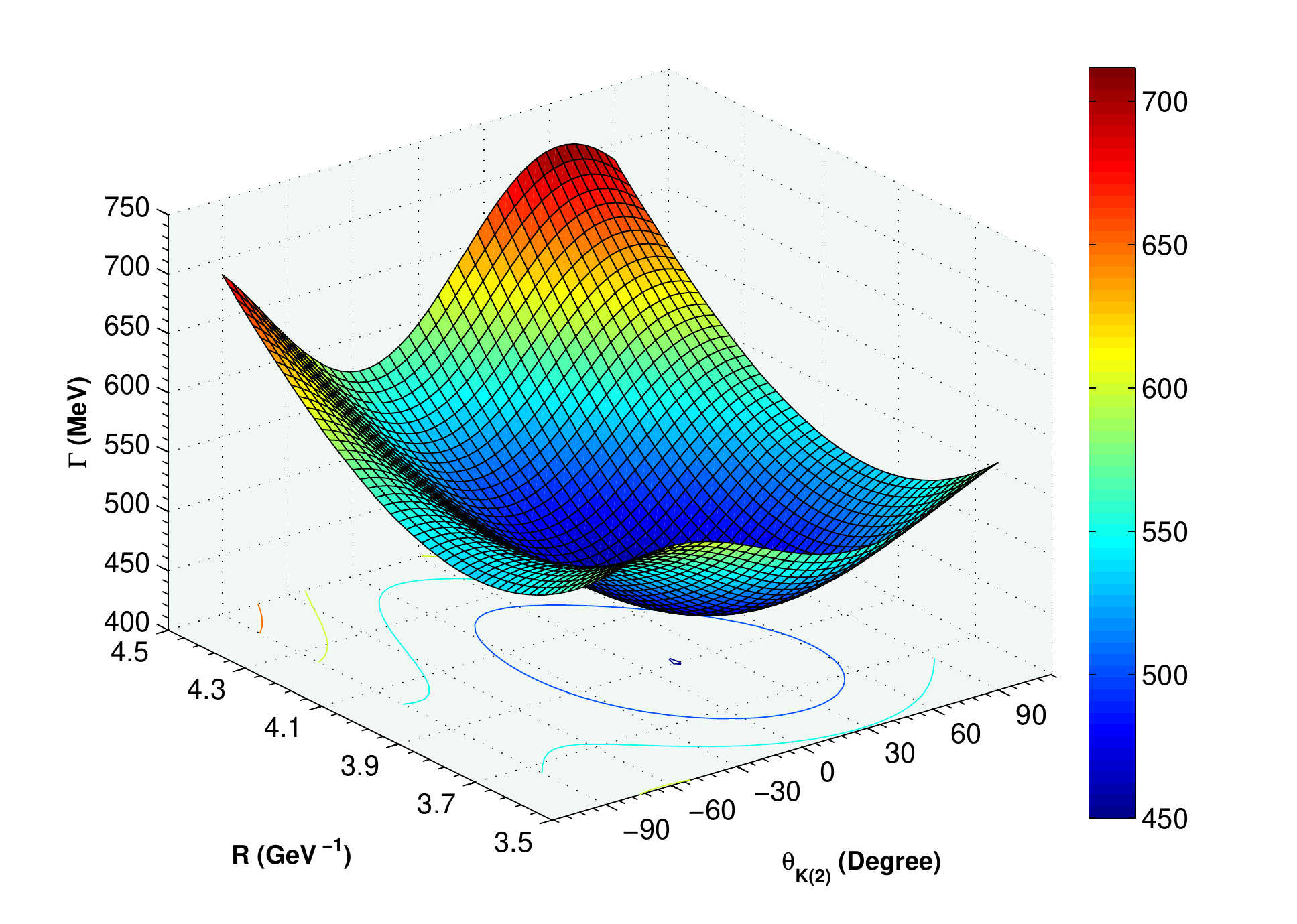}}
\end{minipage}
\hfill
\begin{minipage}{0.45\linewidth}
  \centerline{\includegraphics[width=9cm]{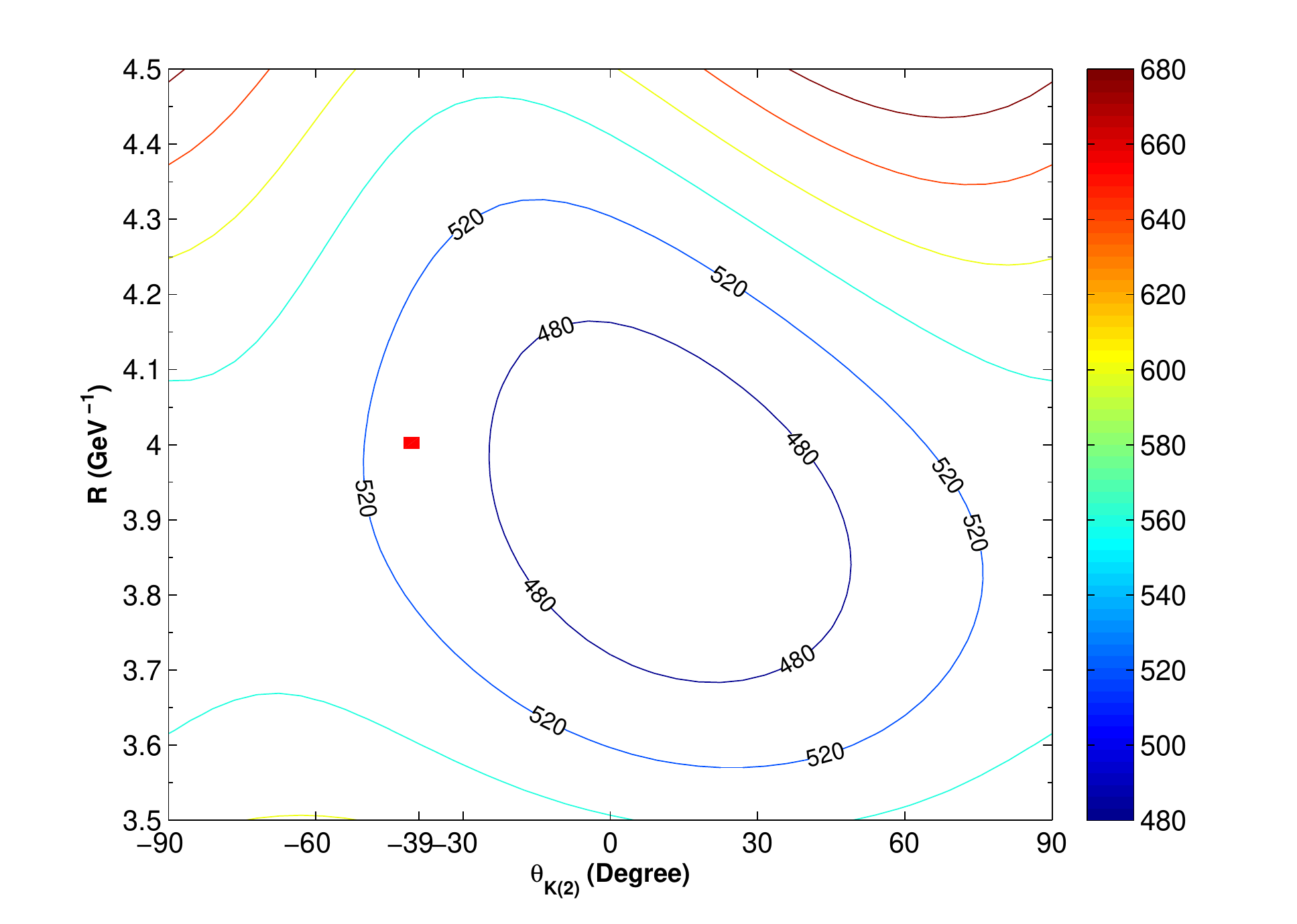}}
\end{minipage}
\caption{(color online). The {three dimensional} (left panel) and contour (right panel) plots of the total decay width of $K_2(2200)$ (the first row), $K_2(2250)$ (the second row) depending on the $R$ and $\theta_{K(2)}$ values.  \label{K2Figure1}}
\end{figure*}

\begin{table}[htbp]
\caption{The typical partial decay widths of $K_2(2200)$ and $K_2(2250)$ (in units of MeV) when taking typical $R=4.0$ GeV$^{-1}$ and $\theta_{K(2)}=-39^\circ$. \label{K2Table1}}
\begin{center}
 \renewcommand{\arraystretch}{1.5}
 \tabcolsep=1.4pt
\begin{tabular}{lcc|lcc}
\toprule[1pt]
\toprule[1pt]
Channels&                            $K_2(2200)$&        $K_2(2250)$&       Channels&        $K_2(2200)$&       $K_2(2250)$\\
\midrule[1pt]

$a_{0}(980)K$&                      0.9&                    10.2&       $K^{\ast}\rho$&        22.2&                19.4\\

$a_{1}(1260)K$&                     1.2&                   11.2&        $K^{\ast}\omega$&      7.5&                 6.5\\

$b_{1}(1235)K$&                     1.8&                   4.4&       $K_1(1270)\pi$&        18.9&                   3.5\\

$h_{1}(1170)K$&                     1.2&                   2.7&        $K_1(1400)\pi$&        2.9&                 3.3\\

$\rho K$&                           51.7&                   7.9&       $K^{\ast}(1410)\pi$&   41.2&                   116.4\\

$\omega K$&                         16.7&                   2.8&       $K_{0}^{\ast}(1430)\pi$&   10.0&               0.009\\

$\phi K$&                           7.5&                   3.6&       $K_{2}^{\ast}(1430)\pi$&   39.3&              14.1\\

$K^{\ast}\eta$&                     7.7&                   7.5&       $f_{2}(1270)K$&            6.3&                13.8\\

$K^{\ast}\pi$&                      20.1&                   71.6&       $f_{1}(1285)K$&            0.9&              3.5\\

$a_1(1260)K^{\ast}$&                25.0&                   19.3&         $b_1(1235)K^{\ast}$&       21.9&            32.3\\

$f_2(1270)K^{\ast}$&                0.5&                   8.6&         $f_1(1285)K^{\ast}$&       3.3&             2.6\\

$h_1(1170)K^{\ast}$&                10.3&                   15.6&         $\rho(1450)K$&            38.6&            40\\

$f_2^\prime(1525)K$&                6.1&                    0.87&         $K^{\ast}\eta^{\prime}$&    3.27&            4.0\\

$K_1(1270)\rho$&                    54.6&                   29.8&         $K_1(1270)\omega$&          17.3&           9.5\\

$K^{\ast}(1410)\rho$&               3.8&                   28.8&         $K^{\ast}(1680)\pi$&        9.5&            16.3\\
\bottomrule[1pt]
\bottomrule[1pt]
\end{tabular}
\end{center}
\end{table}

\subsubsection{The predicted $K_{2}(2560)$ and $K_{2}(2610)$}

\begin{figure*}[htbp]
\begin{minipage}{0.45\linewidth}
  \centerline{\includegraphics[width=9cm]{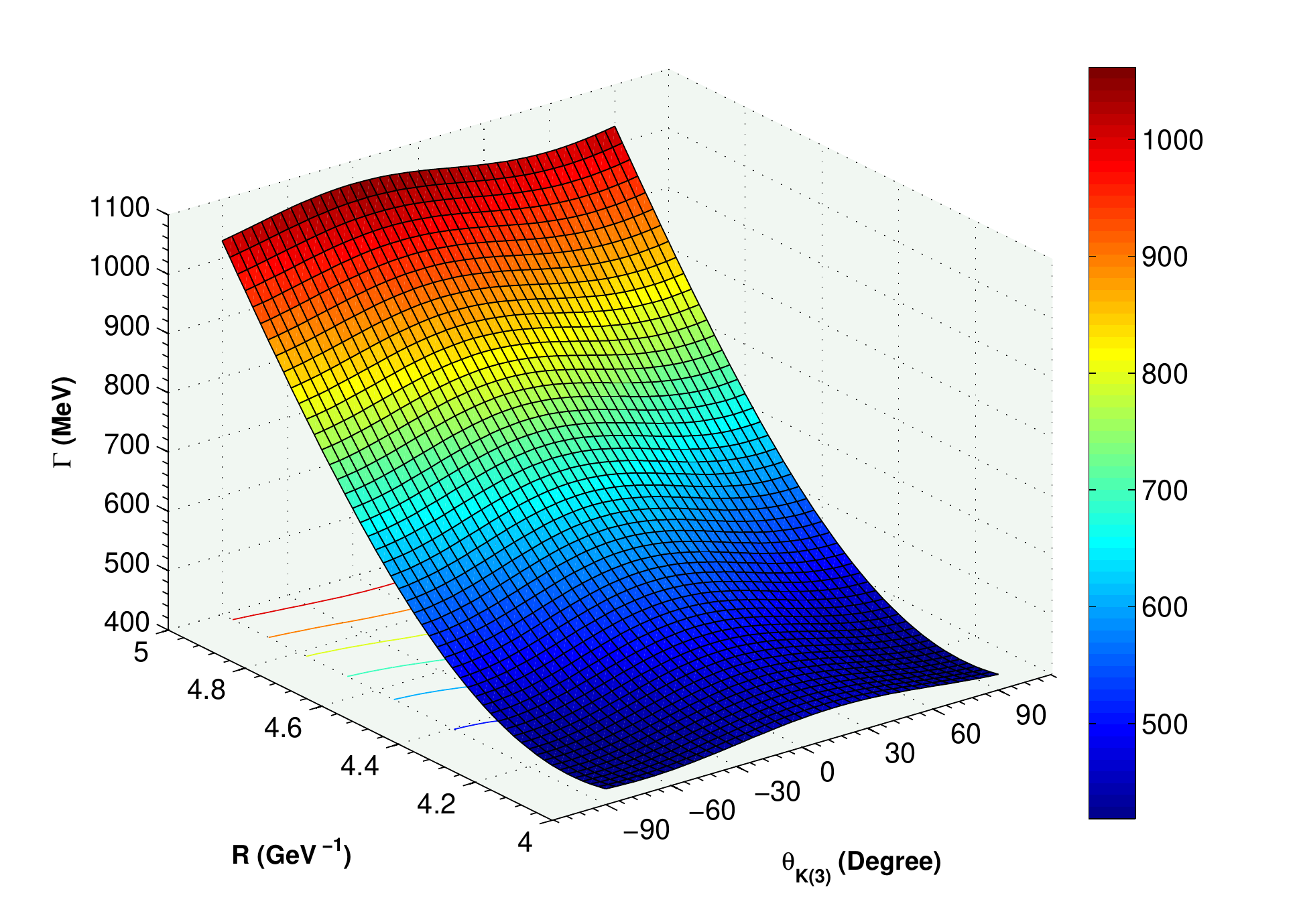}}
\end{minipage}
\hfill
\begin{minipage}{0.45\linewidth}
  \centerline{\includegraphics[width=9cm]{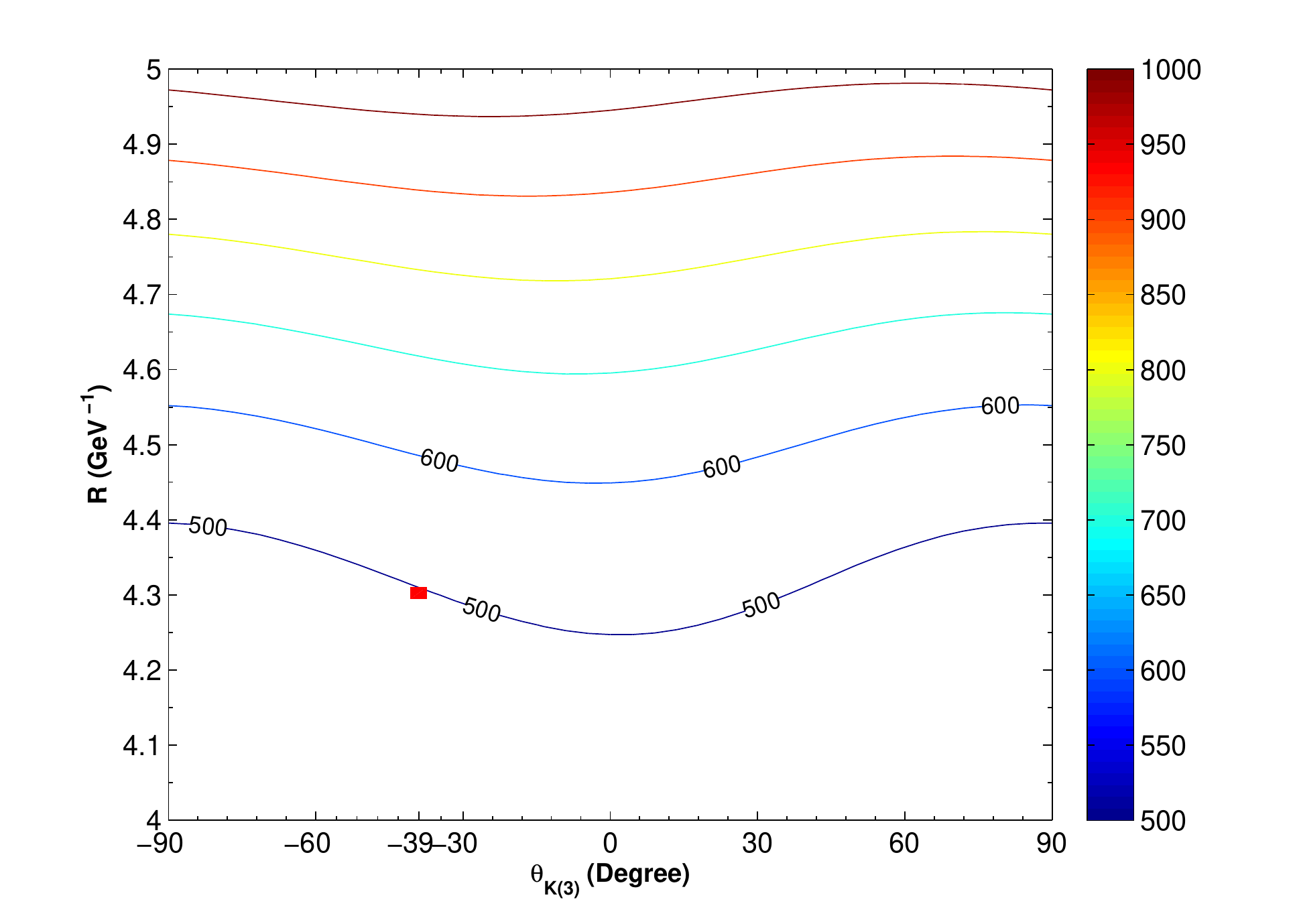}}
\end{minipage}
\vfill
\begin{minipage}{0.45\linewidth}
  \centerline{\includegraphics[width=9cm]{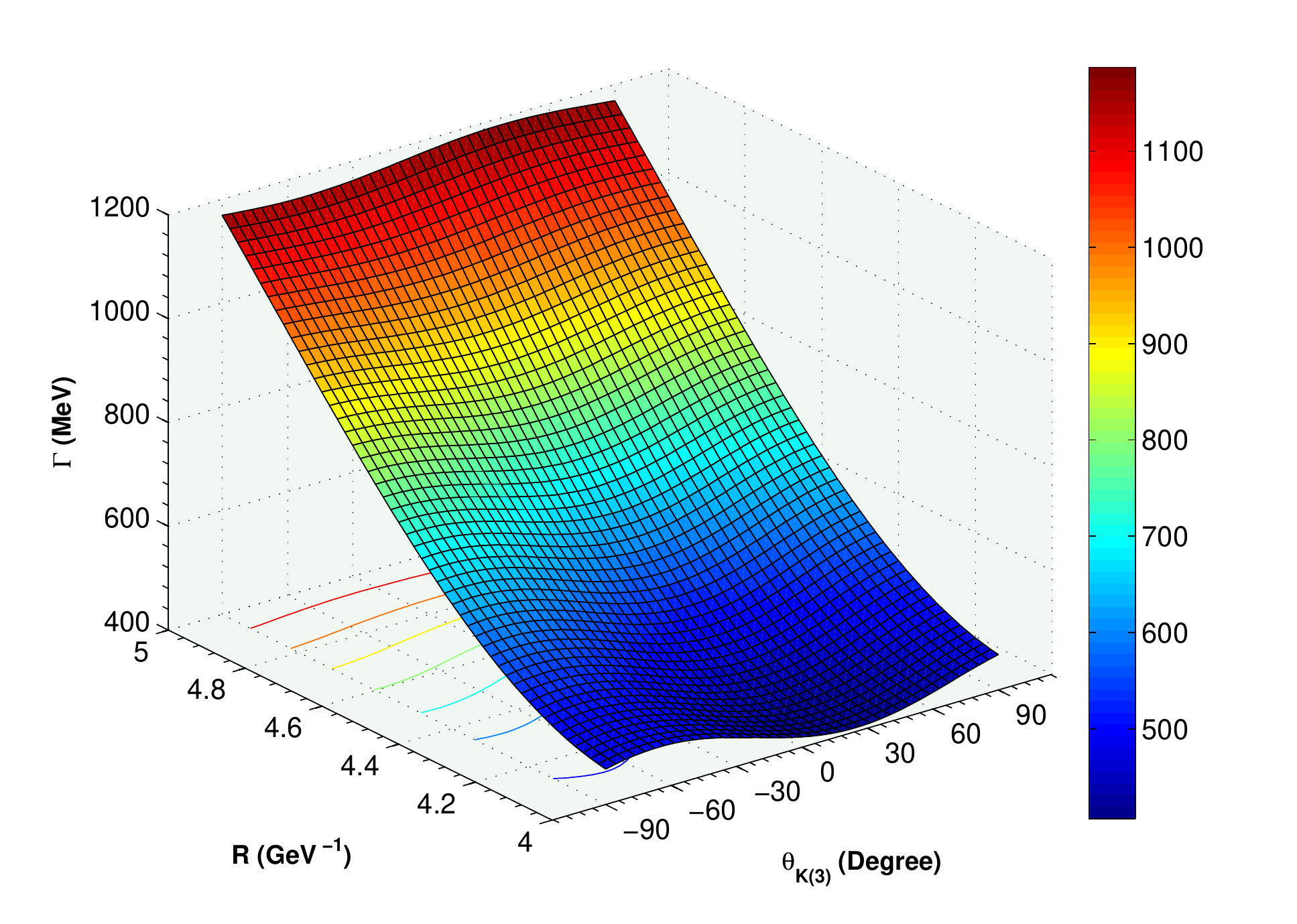}}
\end{minipage}
\hfill
\begin{minipage}{0.45\linewidth}
  \centerline{\includegraphics[width=9cm]{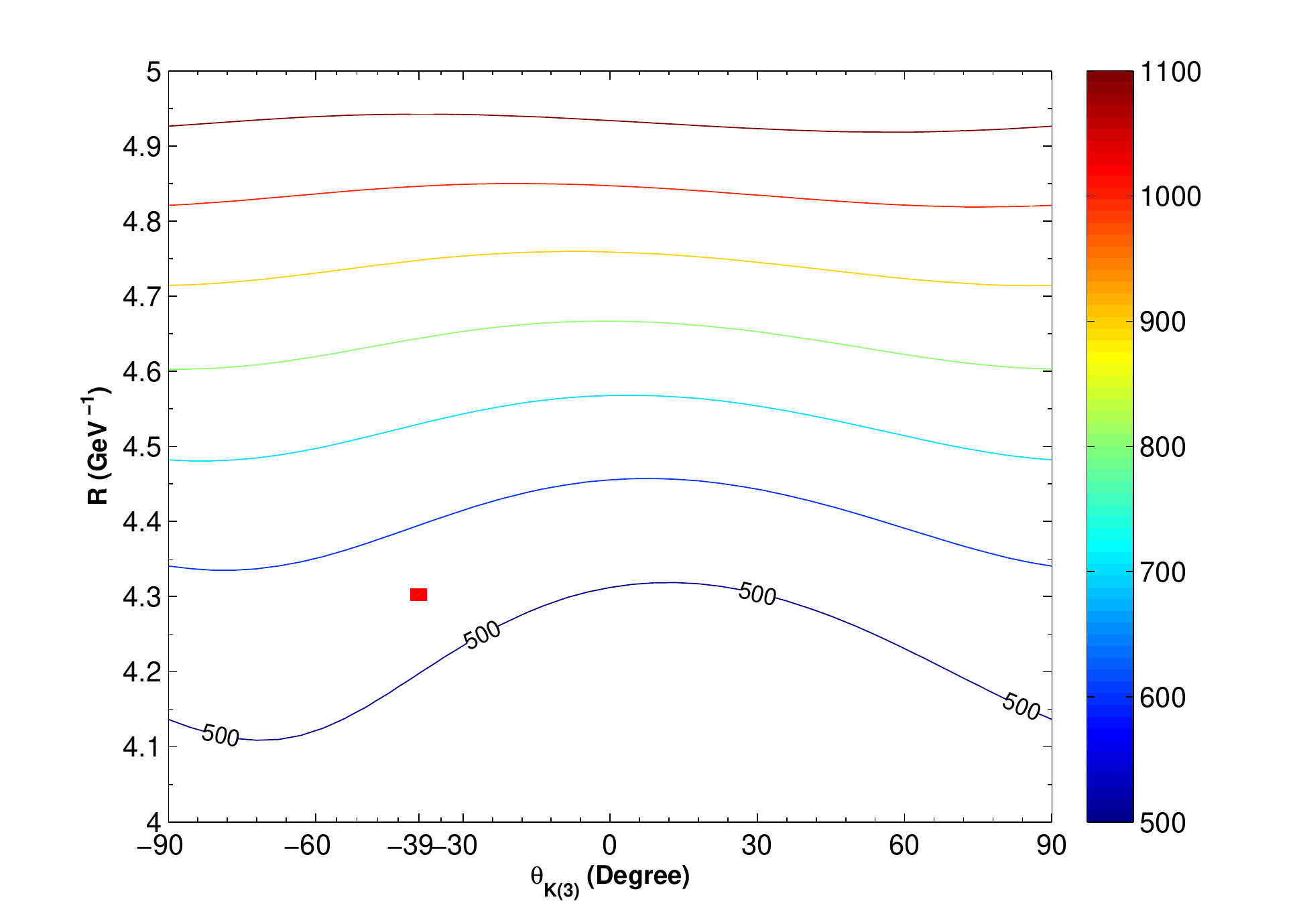}}
\end{minipage}
\caption{The {three dimensional} (left panel) and contour (right panel) plots of the total decay widths of $K_2(2560)$ (the first row), $K_2(2610)$ (the second row) depending on the $R$ and $\theta_{K(3)}$ values. \label{K2Figure2}}
\end{figure*}

\begin{table}[htb]
\caption{The typical partial decay widths of $K_2(2560)$ and $K_2(2610)$ (in units of MeV) when taking typical $R=4.3$ GeV$^{-1}$ and $\theta_{K(2)}=-39^\circ$.\label{K2Table2}}
\begin{center}
 \renewcommand{\arraystretch}{1.5}
 \tabcolsep=1.4pt
\begin{tabular}{lcc|lcc}
\toprule[1pt]
\toprule[1pt]
Channels&                            $K_2(2560)$&        $K_2(2610)$&       Channels&        $K_2(2560)$&       $K_2(2610)$\\
\midrule[1pt]

$a_{0}(980)K$&                      6.0&                    11.9&       $K^{\ast}\rho$&        20.5&                36.4\\

$a_{1}(1260)K$&                     3.3&                   7.9&        $K^{\ast}\omega$&    6.5&                 11.6\\

$b_{1}(1235)K$&                     7.3&                   13.9&       $K_1(1270)\pi$&         28.5&               3.4\\

$h_{1}(1170)K$&                     3.5&                   6.5&        $K_1(1400)\pi$&       1.6&                 19.1\\

$\rho K$&                           97.6&                   2.2&       $K^{\ast}(1410)\pi$&    21.0&                85.1\\

$\omega K$&                         31.8&                   0.64&       $K_{0}^{\ast}(1430)\pi$&   12.3&             2.1\\

$\phi K$&                           3.0&                   7.4&       $K_{2}^{\ast}(1430)\pi$&   6.2&              21.8\\

$K^{\ast}\eta$&                     5.6&                   13.8&       $f_{2}(1270)K$&            6.5&             1.0\\

$K^{\ast}\pi$&                      3.6&                   112.0&       $f_{1}(1285)K$&           0.08&            1.3\\

$a_1(1260)K^{\ast}$&                18.4&                   10.6&         $b_1(1235)K^{\ast}$&     26.5&            8.5\\

$f_2(1270)K^{\ast}$&                1.6&                    2.4&         $f_1(1285)K^{\ast}$&     4.1&             3.0\\

$h_1(1170)K^{\ast}$&                10.2&                   3.4&         $\rho(1450)K$&          29.6&            13.5\\

$f_2^\prime(1525) K$&               3.1&                    0.86&         $K^{\ast}\eta^{\prime}$&  2.4&            4.0\\

$K_1(1270)\rho$&                    14.7&                   28.0&         $K_1(1270)\omega$&         5.0&          9.3\\

$K^{\ast}(1410)\rho$&               20.0&                   14.0&         $K^{\ast}(1680)\pi$&       21.1&          20.7\\

$\rho(1450)K^{\ast}$&               28.7&                   26.1&           $K^{\ast}(1680)\rho$&       33.1&       55.5\\

$K_2(1770)\pi$&                     13.3&                   22.6&           $K_2(1820)\pi$&            11.8&       1.3\\

$a_2(1320)K^{\ast}$&                4.9&                   7.5&            $a_2(1320)K$&              14.3&       3.1\\
\bottomrule[1pt]
\bottomrule[1pt]
\end{tabular}
\end{center}
\end{table}

As $K_2$ mesons $K_{2}(2560)$ and $K_{2}(2610)$  have the radial quantum number $n=3$ and are still missing in experiment. Thus, their decay information is important for future experimental search for them. $K_{2}(2560)$ and $K_{2}(2610)$ satisfy the relation in Eq. (\ref{SpinMixingEquation3}).
We discuss the results by varying $\theta_{K(3)}$. In addition, we set $R=(4- 5)$ GeV$^{-1}$ because of
the fact that $\pi_2(2285)$ and $\eta_2(2250)$ with the predicted $K_{2}(2560)$ and $K_{2}(2610)$ can be categorized into the same nonet, and these mesons have the similar range of $R$.

In Fig. \ref{K2Figure2}, we list the 3-D and contour plots of the total decay widths of $K_2(2560)$ and $K_2(2610)$ that are dependent on the $R$ and $\theta_{K(3)}$ values, which show that both $K_2(2560)$ and $K_2(2610)$ are very broad states and are difficult to be observed in experiment. In addition, the total decay widths of $K_2(2560)$ and $K_2(2610)$ are not strongly dependent on the mixing angle $\theta_{K(3)}$.

To quantitatively illustrate the information of their partial decay widths, Table \ref{K2Table2} includes typical partial decay widths of $K_2(2560)$ and $K_2(2610)$ (in units of MeV) when taking typical $R=4.3$ GeV$^{-1}$ and $\theta_{K(2)}=-39^\circ$. Here, the main decay channels of $K_2(2560)$ are $\rho K$, $K^*(1680)\rho$, $\omega K$, $\rho(1450)K$, $\rho(1450)K^*$, $K^*(1680)\pi$, $K_1(1270)\pi$, $K^*\rho$, $b_1(1235)K^*$, $K^*(1410)\pi$, and $K^*(1410)\rho$. On the other hand, the main decay channels of $K_2(2610)$ are $K^*\pi$, $K^*(1410)\pi$, $K^*(1680)\rho$, $K^*\rho$, $K_1(1270)\rho$, $\rho(1450)K^*$, $K_2^*(1430)\pi$, and $K^*(1680)\pi$. Although $K_2(2560)$ and $K_2(2610)$ are broad resonances according to our calculation, we still suggest the experiments to search for them.
It is obvious that these predicted partial decays can {provide} crucial information for {the experimental study of} these two missing states.

\section{Summary}\label{section5}

In PDG \cite{Beringer:1900zz}, there are abundant observed pseudotensor states with spin-parity quantum number $J^P=2^-$. Inspired by the present experimental status, we carry out the systematical study of the pseudotensor meson family in this work. By the analysis of the Regge trajectories, we discuss the possible categorization of these observed pseudotensor states into three subfamilies, i.e., the $\pi_2$, $\eta_2$, and $K_2$ meson families. In addition, several higher $\pi_2$, $\eta_2$, and $K_2$ mesons still missing in experiment are predicted.

To test these possible assignments, we further investigate the corresponding partial decay behaviors of the discussed pseudotensor states, where the QPC model is adopted. Our study provides { important information on} their main and subordinate decay channels, which is valuable for further experimental investigation {of} these observed states and future {searches} for these predicted higher pseudotensor mesons.

In summary, the studies presented in this work focus on abundant observed pseudotensor states. Our work is helpful to establish the pseudotensor meson family. In addition,
we expect that our work can stimulate experimentalists' interest in exploring higher pseudotensor mesons. Since the main physical aims of COMPASS, BESIII, and forthcoming PANDA {experiments} include the study of light hadrons, these facilities will be a potential and good platform { for exploring} the pseudotensor states.

\hfil
\section*{Acknowledgments}

We would like to thank David Bugg for {the} useful  discussion. This project is supported by the National Natural Science
Foundation of China under Grants No. 11222547, No. 11175073, and No. 11035006, the Ministry of Education of China (SRFDP under Grant No. 2012021111000), and the Fok Ying Tung Education Foundation
(Grant No. 131006).

\section*{Appendix: DEDUCTION OF EQ. (\ref{hhh})}

{The partial wave basis $|J_A, M_{J_A}, J, L\rangle$ is related to the helicity basis $|J_A, M_{J_A}, \lambda_B, \lambda_C\rangle$ through the Jacob-Wick formula \cite{Jacob:1959at},
\begin{eqnarray}
\label{a1}
|J_A, M_{J_A}, J, L\rangle&=&\sum_{\lambda_B, \lambda_C}\sqrt{\frac{2L+1}{2J_A+1}}\langle L0;J\lambda|J_A\lambda\rangle \nonumber \\
&\times& \langle J_B\lambda_B;J_C-\lambda_C|J\lambda\rangle|J_A, M_{J_A}, \lambda_B, \lambda_C\rangle,
\end{eqnarray}
where $\lambda_B$ and $\lambda_C$ are the helicities of final states $B$ and $C$, respectively, and $\lambda=\lambda_B-\lambda_C$. Thus, in the rest frame of the initial state $A$, the partial wave amplitude can be expressed as
\begin{eqnarray}
&&\mathcal{M}^{JL}(A\to BC)(P)\nonumber\\&&=\langle J_A,M_{J_A},J,L|\mathcal{T}|J_AM_{J_A}\rangle\nonumber\\
&&=\sum_{\lambda_B, \lambda_C}\sqrt{\frac{2L+1}{2J_A+1}}\langle L0;J\lambda|J_A\lambda\rangle \nonumber \\&&\quad\times\langle J_B\lambda_B;J_C-\lambda_C|J\lambda\rangle\langle J_A, M_{J_A}, \lambda_B, \lambda_C|\mathcal{T}|J_AM_{J_A}\rangle,
\end{eqnarray}
in which
\begin{eqnarray}
&&\langle J_A, M_{J_A}, \lambda_B, \lambda_C|\mathcal{T}|J_AM_{J_A}\rangle\nonumber\\
&=&\sqrt{\frac{2J_A+1}{4\pi}}\int d\Omega D_{M_{J_A}\lambda}^{J_A}(\phi,\theta,0)\langle\Omega, \lambda_B, \lambda_C|\mathcal{T}|J_AM_{J_A}\rangle\nonumber\\
&=&\sqrt{\frac{2J_A+1}{4\pi}}\int d\Omega D_{M_{J_A}\lambda}^{J_A}(\phi,\theta,0)\nonumber\\
&&\times\langle0, 0, \lambda_B, \lambda_C|U^{-1}[R]\mathcal{T}U[R]U^{-1}[R]|J_AM_{J_A}\rangle\nonumber\\
&=&\sqrt{\frac{2J_A+1}{4\pi}}\int d\Omega D_{M_{J_A}\lambda}^{J_A}(\phi,\theta,0)\sum_{M^{\prime}}D_{M_{J_A}M^{\prime}}^{J_A*}(\phi,\theta,0)\nonumber\\
&&\times\langle0, 0, \lambda_B, \lambda_C|\mathcal{T}|J_A M^{\prime}\rangle\nonumber\\
&=&\sum_{M^{\prime}}\sqrt{\frac{2J_A+1}{4\pi}}\frac{4\pi}{2J_A+1}\delta_{\lambda M^{\prime}}\langle0, 0, \lambda_B, \lambda_C|\mathcal{T}|J_A M^{\prime}\rangle,
\end{eqnarray}
where $D_{M_{J_A}\lambda}^{J_A}$ is the rotation matrix and $U[R]$ is the unitary operator representing a rotation $R(\phi,\theta,0)$. Then, we choose the direction of $\mathbf{P}$ to lie along the positive $z$ axis , i.e., the momentum direction of the final state $B$. Thus, we have $\lambda_B=M_{J_B}$, $\lambda_C=-M_{J_C}$, $|0,0, M_{J_B},M_{J_C}\rangle=\sqrt{4\pi}|0,0, \lambda_B,-\lambda_C\rangle$ and
$\mathcal{M}^{M_{J_A}M_{J_B}M_{J_C}}=\langle0, 0, M_{J_B}, M_{J_C}|\mathcal{T}|J_AM_{J_A}\rangle$.
Thus, one obtains the transformation between the $M^{JL}$ and $M^{M_{J_{A}}M_{J_B}M_{J_C}}$ amplitudes, which reads
\begin{eqnarray}
&&\mathcal{M}^{JL}(A\to BC)=\frac{\sqrt{2L+1}}{2J_A+1}\sum_{M_{J_B},M_{J_C}}\langle L0;JM_{J_A}|J_AM_{J_A}\rangle\nonumber\\
&&\quad\quad\quad\quad\times\langle J_BM_{J_B};J_CM_{J_C}|JM_{J_A}\rangle\mathcal{M}^{M_{J_A}M_{J_B}M_{J_C}},\label{kk1}
\end{eqnarray}
where $M_{J_A}=M_{J_B}+M_{J_C}$. We need to emphasize that in our calculation the factor $\sqrt{4\pi}$
is included into $\mathcal{M}^{M_{J_A}M_{J_B}M_{J_C}}$. Thus, there is no factor  $\sqrt{4\pi}$ appearing in the {rhs} of Eq. (\ref{hhh}).
Finally, the relation listed in Eq. (\ref{hhh}) is obtained. }

\vfil

\end{document}